\documentclass[]{emulateapj}
\usepackage{apjfonts}
\newcommand{\be}{\begin{equation}}
\newcommand{\ee}{\end{equation}}
\newcommand{\bea}{\begin{eqnarray}}
\newcommand{\eea}{\end{eqnarray}}
\newcommand{\Msun}{M_{\odot}}
\newcommand{\comment}[1]{}

\shortauthors{CONROY, SCHIMINOVICH, \& BLANTON}
\shorttitle{EVIDENCE FOR THE 2175\AA\, DUST FEATURE}

\begin{document}
\journalinfo{The Astrophysical Journal}
\submitted{Submitted to the Astrophysical Journal}

\title{Dust attenuation in disk--dominated galaxies: evidence for the
  2175\AA\, dust feature}

\author{Charlie Conroy\altaffilmark{1}, David
  Schiminovich\altaffilmark{2}, \& Michael R. Blanton\altaffilmark{3}}

\altaffiltext{1}{Department of Astrophysical Sciences, Princeton
  University, Princeton, NJ, USA}
\altaffiltext{2}{Department of Astronomy, Columbia University, New
  York, NY, USA}
\altaffiltext{3}{Center for Cosmology and Particle Physics, New York
  University, New York, NY, USA}

\begin{abstract}

  The attenuation of starlight by interstellar dust is investigated in
  a sample of low redshift, disk--dominated star--forming galaxies
  using photometry from {\it GALEX} and SDSS.  By considering
  broadband colors as a function of galaxy inclination we are able to
  confidently separate trends arising from increasing dust opacity
  from possible differences in stellar populations, since stellar
  populations do not correlate with inclination.  We are thus able to
  make firm statements regarding the wavelength--dependent attenuation
  of starlight for disk--dominated galaxies as a function of
  gas--phase metallicity and stellar mass.  All commonly employed dust
  attenuation curves (such as the Calzetti curve for starbursts, or a
  power-law curve) provide poor fits to the ultraviolet colors for
  moderately and highly inclined galaxies.  This conclusion rests on
  the fact that the average FUV-NUV color varies little from face-on
  to edge-on galaxies, while other colors such as NUV$-u$ and $u-r$
  vary strongly with inclination.  After considering a number of model
  variations, we are led to speculate that the presence of the strong
  dust extinction feature at 2175\AA\, seen in the Milky Way (MW)
  extinction curve is responsible for the observed trends.  If the
  2175\AA\, feature is responsible, these results would constitute the
  first detection of the feature in the attenuation curves of galaxies
  at low redshift.  Independent of our interpretation, these results
  imply that the modeling of dust attenuation in the ultraviolet is
  significantly more complicated than traditionally assumed.  These
  results also imply a very weak dependence of the FUV-NUV color on
  total FUV attenuation, and we conclude from this that it is
  extremely difficult to use only the observed UV spectral slope to
  infer the total UV dust attenuation, as is commonly done.  We
  propose several simple tests that might finally identify the grain
  population responsible for the 2175\AA\, feature.

\end{abstract}

\keywords{dust, extinction --- ultraviolet: galaxies --- galaxies:
  ISM}


\section{Introduction}
\label{s:intro}

Dust is everywhere.  Its obscuring and emissive signatures have been
observed in environs ranging from nearby dwarf irregulars to giant
ellipticals, within our own and nearby galaxies, in intergalactic
space, and even in the highest redshift quasars
\citep[e.g.,][]{Trumpler30, Disney89, Knapp89, Goudfrooij94,
  Zaritsky94b, Xu95, Wang96, Elbaz99, Xilouris99, Calzetti01, Motta02,
  Wang08, Menard09}.

Our understanding of the composition and spatial distribution of dust
in a variety of environments has increased markedly in the past
decade, thanks largely to the launches of the {\it Spitzer Space
  Telescope} and the {\it Galaxy Evolution Explorer (GALEX)}.  For
example, the fraction of a galaxy's gas mass that is locked up in dust
is now well-characterized as a function of galaxy mass, star formation
rate (SFR), and metallicity \citep[e.g.,][]{Draine07, daCunha08,
  daCunha10}.  The radial distribution of gas and dust has also been
extensively studied \citep{Munoz-Mateos09}.  The importance and
prevalence of polycyclic aromatic hydrocarbons (PAHs) is now widely
acknowledged thanks to moderate resolution spectroscopy in the mid--IR
\citep[e.g.,][]{Leger84, Smith07, ODowd09}.

Despite these advances, basic questions remain.  It is still not
known, for example, if the majority of dust forms in AGB outflows, in
supernovae ejecta, or condenses directly out of the ISM
\citep{Galliano08, Draine09}.  Computation of the dust extinction
curve from first principles relies on a number of uncertain inputs,
including the size, shape, and composition of the grains, and the
optical constants of the constituent materials
\citep[e.g.,][]{Weingartner01, Draine03, Gordon03, Zubko04}.
A further uncertainty with broad implications for our understanding of
galaxies is the relation between the stellar and dust content of
galaxies; for example, how the grain size distribution varies within
and between galaxies.  Due to gas inflow and outflow, the stellar and
gas--phase metallicities are not simply related, further compounding
the problem.  Without a solid theoretical understanding of the complex
relation between stellar and dust content, these two components need
to be modeled flexibly and independently.  Despite this widely
acknowledged fact, in practice a particular dust obscuration model is
often simply assumed in order to then infer the underlying stellar
population.  Such an assumption can introduce significant biases, as
we will discuss below.

The net effect of dust obscuration within a galaxy is the result of
the combination of the underlying dust extinction curve, geometry, and
radiative transfer.  Here, geometry is taken to mean not only the
large--scale distribution of dust with respect to the stars, but also
the local geometry, which includes the clumpiness of the interstellar
medium (ISM).  Geometrical effects can make the relation between the
underlying extinction curve and the resulting attenuation curve (which
measures the net loss of photons within a galaxy) arbitrarily
complicated \citep[e.g.,][]{Natta84, Witt92, Calzetti94, Witt96,
  Varosi99, Witt00, Granato00, Charlot00, Pierini04, Tuffs04,
  Panuzzo07}.  For example, a clumpy ISM, arising for example from
turbulence, will result in an attenuation curve that is greyer than
the extinction curve.

When the optical depth approaches unity, the effects of geometry
become especially important.  In such cases, the light measured in the
blue/ultraviolet may not trace the same regions of the galaxy as the
red/near--IR light.  Consider a disk galaxy viewed edge--on that is
optically thick in the UV and optically thin in the near-IR.  The
ultraviolet light emitted from the far side of the galaxy will be
heavily extinguished, and so the ultraviolet light collected by an
observer will be preferentially sampling the near side of the galaxy.
In contrast, the near--IR will more faithfully trace the entire
stellar population because the galaxy is relatively transparent in
this wavelength range.  These effects are rarely considered when
modeling the SEDs of galaxies.

If we knew the underlying, unobscured stellar population, the ratio
between the intrinsic and observed light would then provide a direct
measure of the attenuation.  Unfortunately, it is not possible to know
unambiguously the underlying stellar population.  Less direct methods
are therefore required.  For example, the attenuation curve can be
estimated in a sample of galaxies with similar stellar populations and
variable amounts of dust.  In such cases, ratios between more and less
heavily attenuated spectra can provide estimates of the net
attenuation curve in the sample.  Such a technique was utilized by
\citet{Calzetti94} in order to probe the average attenuation curve of
starburst galaxies.  The Balmer decrement was used as an estimator for
the amount of dust attenuation, and, with the assumption that the
galaxies in their sample were of similar metallicity and SFR, the
average attenuation properties were estimated.

Similar techniques for probing the wavelength--dependent attenuation
have recently been applied by \citet{Johnson07b, Johnson07a} to
photometry of a large sample of low redshift galaxies,
\citet{Conroy10b} who analyzed the restframe UV photometry of
star--forming galaxies at $z\sim1$, and \citet{Noll09} who analyzed
spectra of star--forming galaxies at $z\sim2$.  In the present work we
follow in this vein by considering the colors of disk--dominated
galaxies as a function of their inclination.  Since inclination will
correlate with dust attenuation but not with stellar populations,
comparing less to more inclined systems allows us to isolate the
effects of increasing dust opacity on the observed properties of
galaxies \citep[e.g.,][]{Giovanelli94, Giovanelli95, Masters03,
  Driver07, Unterborn08, Maller09, Masters10, Yip10}.

Another constraint on the attenuation curve comes from the relation
between the ratio of total infrared to UV luminosity and UV spectral
slope ($L_{\rm TIR}/L_{\rm UV}-\beta$, or the `IRX--$\beta$'
relation).  Star-forming galaxies with redder UV spectra tend to have
higher IRX.  It is widely believed that this relation is primarily a
sequence in dust attenuation \citep[e.g.,][]{Kong04}.  Various recipes
have thus been proposed to use the IRX--$\beta$ relation to estimate
the total UV attenuation based on the observed UV spectral slope
\citep[e.g.,][]{Buat05, Burgarella05, Cortese08}.  This relation is of
particular importance to the study of high--redshift galaxies, where
only restframe UV and optical photons can be readily collected for
large samples of galaxies \citep[although significant samples at
$z\sim1-2$ with restframe IR data are growing rapidly,
e.g.,][]{Reddy06a, Salim09, Reddy10}.  For most studies of high
redshift galaxies, the observed UV slope, $\beta$, is used in
conjunction with a locally estimated IRX--$\beta$ relation to estimate
the dust opacity.  This approach is essential, for example, to
interpret recent observations of galaxies at $z\approx6-8$
\citep[e.g.,][]{Bouwens09}.

The IRX--$\beta$ relation is different for starbursts \citep{Meurer99}
and normal star--forming galaxies \citep{Dale07, Boissier07}, and in
addition depends somewhat on the star formation history \citep{Kong04,
  Johnson07a, Munoz-Mateos09}, 60$\mu m$ luminosity
\citep{Takeuchi10}, bolometric luminosity \citep{Reddy06a}, and the
star--dust geometry \citep{Panuzzo07}.  These dependencies result in
substantial scatter in the IRX--$\beta$ plane, which, in conjunction
with the fact that the IRX--$\beta$ is nearly vertical over much of
the relevant parameter space (i.e., IRX varies considerably over a
relatively narrow range in $\beta$), calls its utility into question
\citep{Bell02b}.

The most prominent feature of the MW extinction curve is the strong,
broad dust feature at $2175$\AA, the `UV bump' \citep{Stecher65}.
This feature is also seen, albeit more weakly, along most sightlines
in the LMC and along one of the five sightlines probed in the SMC
\citep{Gordon03}.  It is also seen along sightlines in M31
\citep{Bianchi96}.  There are now confident detections of this feature
in the extinction curves of more distant galaxies as well, as probed
by background gamma ray bursts and quasars, and gravitational lenses
\citep{Ardis09, Motta02, Wang04, Mediavilla05}, Puzzlingly, there are
many examples of galaxies that do not show this feature in their
extinction curve \citep[e.g.,][]{York06, Stratta07}.  The carrier of
this extinction feature is not known, although owing to its strength
it must be due to some abundant material, such as carbon.
Measurements of the grain albedo suggest that the feature is not due
to scattering \citep[see data compilations in][]{Witt00, Draine03}.
Several dust models associate this feature with PAH absorption
\citep[e.g.,][]{Weingartner01} although there are other possibilities
\citep[e.g.,]{Draine93}.

Understanding the prevalence of the UV bump in the attenuation curves
of other galaxies is essential for broadband photometric studies of
galaxies.  For example, at $z\sim0$ the UV bump falls into the {\it
  GALEX} NUV-band, at $z\sim1$ it falls within the $B-$band, and at
$z\sim2$ it redshifts into the $R-$band.  If present in the
attenuation curves of galaxies, the UV bump would therefore result in
substantially more, and likely more uncertain, attenuation in
particular filters at particular redshifts.

Based on spectra from the {\it International Ultraviolet Explorer},
Calzetti et al. found an average attenuation curve for starburst
galaxies that lacked a UV bump.  The absence of a UV bump in the
starburst attenuation curve led \citet{Witt00} to suggest that in such
galaxies the underlying dust extinction curve lacks a UV bump.  A
detailed analysis of the ultraviolet through infrared photometry of
M51 revealed little evidence for a UV bump within individual HII
regions \citep{Calzetti05}.  \citet{Conroy10b} also found no evidence
for a UV bump as strong as that seen in the MW in star--forming
galaxies at $z\sim1$.  In stark contrast, \citet{Noll09} presented
strong evidence for a UV bump in stacked restframe UV spectra of
$z\sim2$ star--forming galaxies.  Finally, \citet{Capak10} presented
tantalizing evidence for a strong UV bump in at least one $z\sim7$
galaxy.

Thus, while the UV bump appears to be a ubiquitous feature of the MW
and LMC extinction curves, there is little evidence for this feature
in the net attenuation curves of other galaxies.  However, it is
important to recognize that a systematic investigation of the UV
attenuation in $z\sim0$ `normal' star-forming galaxies is currently
lacking.  While the effects of geometry and scattering can diminish
the strength of the UV bump with respect to the underlying extinction
curve, it is a generic prediction of radiative transfer calculations
that if the UV bump is present in the extinction curve, its presence
will be detectable in the resulting attenuation curve \citep{Witt00,
  Panuzzo07}.  Given the observations of the UV bump in the extinction
curves of other galaxies, the attenuation curves of at least some
galaxies should show a UV bump, even if the strength of the bump is
weak.  This expectation serves as motivation for the present study.

There is currently no clear picture linking together these various
observations.  One possibility is that the radiation field modulates
the strength of the UV bump, as suggested by \citet{Gordon03}.  The
typical radiation field in the SMC and the local starburst galaxy
sample is much harsher than the MW, potentially explaining why a UV
bump is not seen in such systems.  It would be hard to explain the
results of Noll et al. in this context, however, given that their
sample is dominated by high SFR systems.  \citet{Granato00} explained
the result from Calzetti et al. as being due to the fact that the UV
energy production in starbursts is dominated by young stars that are
heavily embedded within molecular clouds.  In their model, attenuation
in starbursts is therefore governed by the wavelength-dependent
fraction of UV photons produced by young stars (whose light is heavily
extinguished by their birth cloud).  These authors predicted that the
attenuation suffered by `normal' star-forming galaxies would show
evidence of a UV bump because in this case significant UV energy is
provided by intermediate age stars that have left their birth clouds.
Such stars will suffer attenuation primary from the diffuse dust,
where significant 2175\AA\, absorption may be expected.  It is not
immediately obvious whether or not this explanation can accommodate
the results from Noll et al.  \citet{Draine07} has recently
demonstrated a deficiency of PAH emission in the infrared in galaxies
with low gas--phase metallicities \citep[see also][]{Engelbracht05,
  Smith07}.  Metallicity may therefore play an important role.

In the present work we investigate the wavelength--dependent
attenuation by dust for a sample of disk--dominated galaxies.  Our
sample is carefully selected to be homogeneous and complete.  We then
consider the ultraviolet, optical, and near--infrared colors as a
function of inclination.  Since inclination will only correlate with
dust attenuation and not physical parameters such as star formation
nor metallicity, the inclination--dependent colors will provide a
robust and sensitive probe of the wavelength--dependent attenuation in
disk--dominated systems.  This technique is therefore similar in
spirit to that of \citet{Calzetti94, Calzetti00}, although for a
sample of `normal' star-forming galaxies, and utilizing photometry
rather than spectroscopy.  Averaging the flux of many galaxies
considerably simplifies the modeling of the underlying stellar
population because the average star formation history (SFH) of many
normal galaxies must be smooth.  By normal here we mean galaxies not
chosen to have special SFHs, such as starburst or post-starburst
galaxies.  We will focus especially on the attenuation properties in
the NUV band, and will therefore be able to make strong statements
regarding the presence of the UV bump in the average attenuation curve
of normal star-forming galaxies.

Where necessary, a flat $\Lambda$CDM cosmology with $(\Omega_m,
\Omega_\Lambda)=(0.30,0.70)$ is adopted, along with a Hubble constant
of $H_0 = 100h$ km s$^{-1}$ Mpc$^{-1}$.  All magnitudes are in the
$AB$ system \citep{Oke83}.  A \citet{Chabrier03} initial stellar mass
function (IMF) is adopted when quoting stellar masses.


\section{Data}
\label{s:data}

Redshifts, coordinates, $ugriz$ photometry, and structural properties
are derived from Sloan Digital Sky Survey \citep[SDSS;][]{York00}
data, as made available through the NYU Value Added Galaxy
Catalog\footnote{\texttt{http://sdss.physics.nyu.edu/vagc/}}
\citep[VAGC;][]{Blanton05}.  We do not simply use the photometry from
the SDSS catalogs, because the automated pipeline often incorrectly
de-blends large galaxies.  In addition, the images themselves are
generally over-subtracted due to the sky estimation procedure, which
uses a median smoothed field in 100'' by 100'' boxes
\citep{Blanton05}.  In order to address these issues, Blanton et
al. (in prep) measure the smooth sky background from heavily masked
versions of each imaging run. In this way, we avoid subtracting away
light from large galaxies.  Blanton et al. (in prep) have implemented
a de-blending pipeline similar to that described by \citet{Lupton01},
but with parameters optimized for brighter and larger objects.  For
small objects these measurements agree well in general with the
original SDSS catalog values; for larger objects (half-light radii
$>10''$) our galaxy measurements tend to be brighter in flux and
larger in size.

We use the `lowz' VAGC sample, which is constructed from the SDSS Data
Release 6 \citep{DR6}, but with the photometric re-reductions
discussed above.  This sample includes all galaxies with redshifts
$z<0.05$.  \citet{Sersic68} profile fits have been made to the radial
light profiles of galaxies in this sample (the profile fits were
corrected for seeing).  We will make use of the Sersic index $n$,
where recall that $n=1$ corresponds to an exponential light profile
and $n=4$ corresponds to a de Vaucouleurs profile.  We will also make
use of the axial ratio, $b/a$, where $b$ and $a$ are the semi-minor
and semi-major axes, respectively. The axial ratio is measured within
a radius containing 90\% of the total light, and will be taken to be a
proxy for the galaxy inclination.  Petrosian magnitudes are used
herein.

Gas--phase oxygen abundances have been estimated from the SDSS spectra
by \citet{Tremonti04}, although these measurements are only available
for galaxies in the SDSS Data Release 4.  When considering galaxies
split according to gas--phase metallicities, we will only use galaxies
that are in the Tremonti et al. catalog.

\begin{figure}[!t]
\begin{center}
\resizebox{3.5in}{!}{\includegraphics{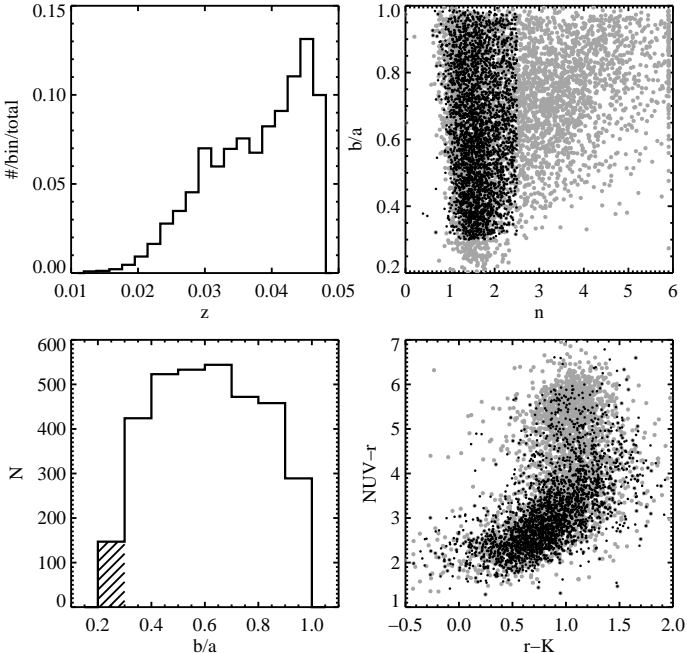}}
\end{center}
\caption{Properties of the low redshift galaxy sample used in the main
  analysis.  {\it Upper left:} Redshift distribution normalized to the
  total number of objects in the sample.  {\it Upper right:} Relation
  between Sersic index $n$ and axial ratio $b/a$ for the full low
  redshift sample (randomly diluted by 60\% for clarity; {\it grey
    symbols}), compared to the subsample used in the main analysis
  ({\it black symbols}).  {\it Lower left:} Histogram of axis ratios
  for the sample used in the main analysis.  Bin widths are 0.1,
  starting at $b/a=0.2$, which corresponds exactly to the bins used in
  later sections.  The hatched bin is excluded from our analysis for
  reasons discussed in the text. {\it Lower right:} Color--color
  diagram (symbols are as in the upper right panel).
  UV--optical--near-IR colors have been shown to efficiently separate
  truly quiescent from dusty star--forming galaxies.  The galaxies
  used in this sample populate the latter locus.}
\label{fig:props}
\vspace{0.2cm}
\end{figure}

Near--IR photometry are provided by the Two Micron All Sky Survey
\citep[2MASS][]{Jarrett00}.  We make use of the $K-$band only in
Figure \ref{fig:props} when investigating the properties of our
sample, and so make no attempt to perform matched aperture photometry
on the 2MASS data.

The {\it GALEX} survey \citep{Martin05} is providing ultraviolet
imaging in the FUV and NUV filters (with effective wavelengths of
1539\AA\, and 2316\AA, respectively).  In order to accurately probe
the UV through optical SED of galaxies as a function of inclination,
we have re-measured {\it GALEX} photometry self-consistently for our
sample by computing magnitudes within the Petrosian aperture defined
by the SDSS photometry (Petrosian magnitudes in SDSS are defined
according to the Petrosian radius, $r_P$, measured in the $r-$band).
Our photometry was performed on {\it GALEX} images that have been
calibrated and background subtracted.  Thus, the FUV, NUV, and $ugriz$
magnitudes are all measured through the exact same aperture.  The
Petrosian aperture (which is defined as $2r_P$) is much larger than
the half--light radius and so differences in the PSF between {\it
  GALEX} and SDSS should have a negligible effect on the derived
photometry, especially for the exponential light profiles that
dominate our sample.  Zero points for the FUV and NUV magnitudes are
adopted from \citet{Morrissey07}.

$E(B-V)$ reddening estimates are derived from \citet{Schlegel98} via
the utilities in the \texttt{kcorrect v4.1.4} software package
\citep{Blanton07}.  Photometry is corrected for Galactic extinction
using $E(B-V)$ and the MW extinction curve of \citet{Cardelli89}.  In
particular, we use the following values for $R_i\equiv A_i/E(B-V)$,
where $A_i$ is the extinction in magnitudes for filter $i$: 8.29,
8.18, 5.16, 3.79, 2.75, 2.09, 1.48, and 0.30 for the FUV, NUV,
$ugriz$, and $K$ bands.

K--corrections are estimated with the routines made available in the
\texttt{kcorrect v4.1.4} package.  The SDSS $ugriz$ photometry is used
as input to estimate k--corrections for both the SDSS photometry and
{\it GALEX} photometry.  Photometry is k--corrected to $z=0.04$, which
is the median redshift of the sample.  Typical k--corrections are of
order 0.01 magnitudes and therefore have no significant effect on our
results; they are included merely for completeness.  

The \texttt{kcorrect v4.1.4} package also provides estimates of the
stellar masses of galaxies given a set of input photometry.  We have
used the SDSS photometry to estimate stellar masses for our sample,
assuming a \citet{Chabrier03} initial mass function.  We prefer to use
photometry for stellar mass estimation because masses derived in this
way are only weakly dependent on galaxy inclination \citep[stellar
masses differ by $<0.1$ dex between highly inclined and face--on
systems;][]{Maller09}.  Other, popular stellar mass estimators based
on spectroscopy \citep{Kauffmann03a, Tojeiro09}, suffer stronger
dependencies with inclination, which indicates a systematic source of
bias in the masses.

Our goal is to explore the inclination--dependent dust characteristics
of a uniform sample of disk--dominated galaxies.  In order to
construct such a sample we select galaxies with $n<2.5$ to separate
disk--dominated from bulge--dominated galaxies.  In addition, galaxies
are selected in a narrow stellar mass range $9.5<{\rm
  log}(M/M_\Sol)<10.0$ in order to define a homogeneous and
volume-limited sample.  Mass and SFR are correlated at $z\sim0$
\citep[e.g.,][]{Brinchmann04, Salim07} and so our selection on mass
may be interpreted roughly as a selection on SFR as well.  Galaxies at
$z<0.01$ are removed from our sample, as they are generally very large
on the sky and thus accurate photometry is complicated.  We also
remove 112 galaxies with $r_P<4"$, as inclination measurements become
unreliable for such small galaxies because of seeing (see below).
With these cuts we are left with 3394 galaxies with detections in all
{\it GALEX} and SDSS filters.  This is our fiducial sample.

As discussed in \citet{Masters03}, the effects of seeing can bias
inclination measurements for galaxies with small angular sizes.  The
bias manifests itself as a deficiency of small, highly inclined
objects.  Indeed, our sample shows such a deficiency for galaxies with
$r_P\lesssim8"$.  The median Petrosian radius of our fiducial sample
is $8".4$, so this bias potentially affects a significant fraction of
our sample with high inclinations.  However, we have re-computed all
of the results in following sections only including galaxies with
$r_P>8"$ (and widening the stellar mass cut to $9.25<{\rm
  log}(M/\Msun)<10.0$ in order to increase the sample size), and find
that the inclination--dependent colors change by an inconsequential
amount.  Moreover, these issues primarily affect galaxies with
$0.2<b/a<0.3$ and since we consider only galaxies with $b/a>0.3$ we
can be confident that our results are robust to seeing biases.

Several properties of our fiducial sample are shown in Figure
\ref{fig:props}, including the redshift distribution, scatter plot of
inclination and Sersic index, distribution of inclinations, and a
color--color plot.  The deficiency of galaxies with $b/a<0.3$ is due
in large part to the seeing bias noted in the previous paragraph.
However, one would also expect the intrinsic number of axial ratios to
drop at such low values because of the rareness of intrinsically thin
disks.  This bin is excluded from our analysis below.

The NUV$-r$ color, in combination with an optical-near-IR color such
as $r-K$, cleanly separates star--forming and quiescent galaxies, even
when there are substantial amounts of dust in star--forming galaxies
\citep{Williams09, Bundy09}.  It is clear from this figure that our
sample is dominated by star--forming galaxies, which is not surprising
given the mass range of the sample.  Galaxies in this mass range have
current SFRs that are comparable to their past--averaged SFRs
\citep[i.e., the birth--parameter for these objects is $\approx
1$;][]{Brinchmann04}.  In other words, star formation has proceeded in
an approximately continuous manner over the lifetime of these
galaxies.  The results presented in following sections are unchanged
if we remove the few truly quiescent galaxies in our sample (galaxies
with NUV$-r$ colors greater than $\approx4.5$).

\subsection{On the challenges and benefits of averaging many galaxies}
\label{s:stack}

In $\S$\ref{s:res} we will present average colors of disk--dominated
galaxies as a function of inclination.  Within each bin in inclination
the average colors are computed by averaging the fluxes for typically
several hundred galaxies.  In the absence of dust, averaging in this
manner is equivalent to constructing a `super-galaxy' that is composed
of all of the stars of the galaxies in the bin.  With the addition of
dust attenuation, the interpretation of this averaging procedure is
more complex.  In this case magnitudes may be somewhat simpler to
interpret because magnitudes are linear in the attenuation.  Indeed,
if the underlying stellar population is unchanged and only the dust
content varies, then averaging magnitudes is probably the correct
approach.  In $\S$\ref{s:res} we will present average colors computed
both by averaging fluxes and averaging magnitudes in order to
demonstrate that our results are insensitive to this distinction.
Unless explicitly stated otherwise, our results will be computed by
averaging fluxes.

The major benefit of averaging many galaxies is that the SFH of the
ensemble must be smooth.  As noted in the introduction, this need not
necessarily be the case if one selects special classes of galaxies
such as starburst or post-starburst galaxies.  Our sample is composed
of normal star-forming galaxies, and so our assertion that the SFH
must be smooth is justified.  One thus cannot appeal to sharp
variations in the SFH (e.g., recent bursts) to explain the average
colors of the stacked galaxies.  This will prove to be an important
asset when interpreting the inclination--dependent colors in
$\S$\ref{s:res}.


\section{Models}
\label{s:models}

\subsection{Stellar population synthesis}

The stellar population synthesis (SPS) treatment closely follows that
of \citet{Conroy09a, Conroy10a}, to which the reader is referred for
details.  Briefly, the code produces the evolution in time of the
spectral energy distribution of a coeval set of stars from $10^{6.5}$
to $10^{10.15}$ yr, for a wide range in metallicities.  Stellar
evolution from the main sequence through the thermally--pulsating AGB
and post--AGB phases are included.  The IMF of \citet{Chabrier03} is
adopted.  

In \citet{Conroy10c}, this SPS code was extensively calibrated against
a suite of observations and was also compared to several other popular
SPS codes.  The version of the SPS code used herein is identical to
that discussed in \citet{Conroy10c}.  We have also compared this SPS
code to the predictions of the Starburst99 code \citep{Leitherer99,
  Vazquez05} for a constant SFH at solar metallicity.  The spectra
from these two models are very similar in the ultraviolet and optical.
This agreement is encouraging because Starburst99 contains a more
sophisticated treatment of massive star evolution and also includes
nebular continuum emission, which is not included in our models.

The underlying stellar population in our default model will consist of
solar metallicity stars that formed in a continuous, constant mode of
star formation from $t=0.0$ to $t=13.7$ Gyr.  In order to explore the
impact of these assumptions we will also consider models with
metallicities $Z=0.5Z_\Sol$ and $Z=1.5Z_\Sol$, and models where a
fraction, $c<1$, of mass forms in a constant mode of star formation,
the rest having formed instantly at $t=0.0$.  All colors considered
herein have been k--corrected to the median redshift of the data,
which is $z=0.04$.

\begin{figure}[!t]
\begin{center}
\resizebox{3.5in}{!}{\includegraphics{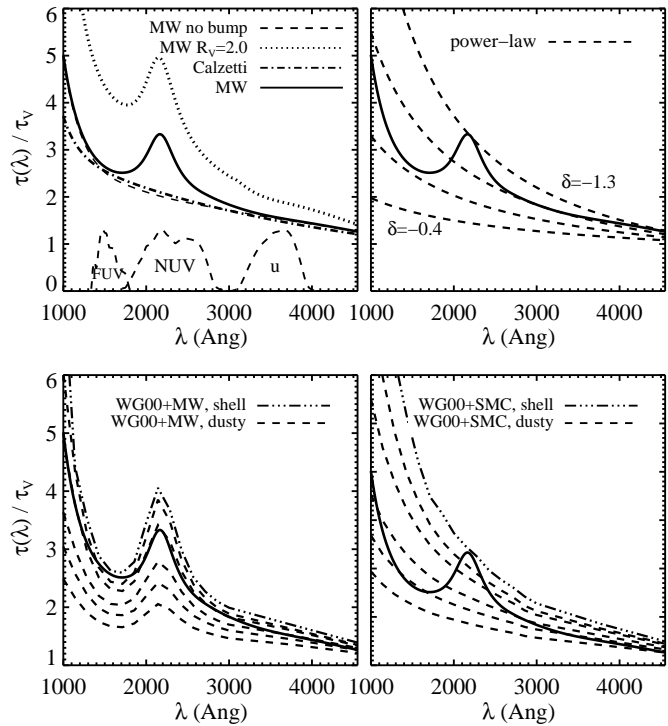}}
\end{center}
\caption{UV attenuation curves, normalized to the attenuation in the
  $V-$band.  The MW curve with $R_V=3.1$ is repeated in all panels for
  reference.  {\it Top left:} The MW curve (with $R_V=3.1$) both with
  and without a UV bump are compared to the Calzetti et
  al. attenuation curve and a MW curve with $R_V=2.0$.  The
  transmission curves for the FUV, NUV, and $u-$band filters are also
  shown.  {\it Top right:} Power--law curves of the form
  $\tau(\lambda)\propto\lambda^\delta$, with $\delta=-0.4$, $-0.7$,
  $-1.0$, and $-1.3$.  {\it Bottom panels:} Attenuation curves derived
  from the models of WG00 for MW and SMC extinction curves are shown
  both for the `shell' and `dusty' geometries.  The $V-$band
  normalized attenuation curve for the dusty geometry depends on the
  dust column density in the sense that larger columns yield grayer
  curves.  In contrast, the shape of the attenuation curve for the
  shell geometry is independent of column density and so only one
  curve is seen for that geometry.}
\label{fig:uvext}
\vspace{0.2cm}
\end{figure}

\vspace{0.5cm}
\subsection{Dust models}
\label{s:dust}

A suite of phenomenological dust models are considered for the
attenuation of starlight.  The models are similar in spirit to the
two--component dust model of \citet{Charlot00}.  This model
distinguishes between attenuation suffered by young stars in their
natal clouds and attenuation of all starlight due to diffuse cirrus
dust.

Attenuation due to diffuse cirrus dust is described by the $V-$band
opacity $\tau_{V,d}\equiv\tau_d$ and an attenuation curve.  By
`attenuation curve' we mean the variation in attenuation optical depth
with wavelength, normalized to 1.0 in the $V-$band, i.e.,
$\tau(\lambda)/\tau_V$.  In this model, stars with ages younger that
$t_{\rm esc}$ are subject to additional attenuation (attributed to
their birth cloud) characterized by an optical depth
$\tau_{V,bc}\equiv\tau_{bc}$ with an attenuation curve that in
principle may differ from the curve describing the diffuse dust.
Based on data from a sample of low redshift star--forming galaxies,
\citet{Charlot00} favor values of $\tau_{bc}=1.0$ and $t_{\rm
  esc}=10^7$ yr with an attenuation curve of the form
$\tau\propto\lambda^{-0.7}$ for both components.  Herein we will
consider additional attenuation curves and model parameters, although
for our default model we will adopt $\tau_{bc}=0.5$ and $t_{\rm
  esc}=10^7$ yr.

This two component model has strong observational motivation not only
from direct observations of young stars embedded in molecular clouds
but also from integrated spectra of star--forming galaxies, where the
opacity measured in balmer emission lines is a factor of roughly two
larger than the opacity measured from the stellar continuum
\citep{Calzetti94}.

We will also briefly consider a dust model that, in addition to the
two component model described above, allows for a `skin' of completely
unobscured starlight, quantified as the fraction, $f$, of the stellar
mass that is unobscured.  Motivation for such a model comes from
recent observations of extended UV emission in the outer edges of many
disk galaxies \citep{Thilker07}.  The young stars in these outer
regions are apparently metal--poor ($Z\sim Z_\Sol/10$), and the
extinction they suffer is small, though non-zero \citep{GildePaz07,
  Werk10}.  In addition, it is well-known that the dust--to--gas ratio
decreases with radius in disk galaxies \citep[e.g.,][]{Issa90,
  Boissier04, Munoz-Mateos09}, and so the outer regions of galaxies
should suffer substantially less extinction than the inner regions.
Finally, a sizeable fraction ($\approx10-30$\%) of OB stars are
`runaways' that have escaped their birth clouds \citep{Stone91}.
These stars will likely still suffer attenuation from the diffuse dust
component, but nonetheless they may constitute another skin-like
population that contributes a bluer than average spectrum to the
integrated light.

\subsubsection{Attenuation curves}
\label{s:attn}

We now summarize the various attenuation curves that are considered
herein.  Simple power--law curves are considered, characterized by an
index $\delta$ such that $\tau\propto\lambda^{\delta}$.  We also
consider attenuation curves whose functional form is equivalent to the
average extinction curve in the MW \citep{Cardelli89}.  As mentioned
in the Introduction, attenuation and extinction are different
concepts, but in simple geometries they can be roughly equivalent
(modulo wavelength dependence of the dust albedo).\footnote{Later,
  when we refer to `MW attenuation' we mean an attenuation curve that
  is equivalent to the observed, average MW extinction curve.} When
considering the MW curves we allow for the freedom to vary the
strength of the UV bump, parameterized by $B$.  In the Appendix we
provide formulae for constructing MW--like extinction curves with
arbitrary UV bump strength.  The default MW curve is recovered for
$B=1.0$; for $B=0.0$ there is no bump.  Below we will consider an
attenuation curve that resembles the MW with $B=1.0$, $B=0.8$, and
$B=0.0$.  We also allow for the freedom to vary the standard
extinction parameter $R_V\equiv A_V/E(B-V)$.  Note that a smaller
$R_V$ corresponds to a steep extinction curve.  The average extinction
curve in the MW is well-fit with $R_V=3.1$, while a value of
$R_V\approx2$ provides a good fit to the average SMC extinction curve
at $\lambda\lesssim5000$\AA\, \citep{Pei92}.  Below, when we refer to
`MW extinction' generically, we mean the standard, average MW curve
with $R_V=3.1$ and $B=1.0$.

Attenuation curves produced by the models of \citet[][WG00]{Witt00}
will also be considered.  These authors generate attenuation curves
from both MW (adopting $R_V=3.1$) and SMC extinction curves, taking
into account the effects of scattering and geometry via radiative
transfer calculations.  They produce curves for three large--scale
star--dust geometries: a `shell' geometry where all of the dust is in
front of the stars (i.e., a uniform screen), a `dusty' geometry where
the stars and dust are equally mixed within a sphere, and a `cloudy'
geometry where stars and dust are equally mixed within the inner 70\%
of the system, with the remaining outer portion containing stars but
no dust.  The WG00 models assume spherical symmetry.  For each of
these large--scale geometries WG00 consider both a homogeneous and
clumpy distribution for the local distribution of dust.  There are
twelve models in all (two extinction curves, three large--scale
geometries, and two local geometries).  Each model is computed for a
wide range in column densities.

The attenuation curves considered herein are shown in Figure
\ref{fig:uvext}.  For the WG00 models, only the homogeneous local
geometries are shown, and only for the shell and dusty large--scale
geometries.  Notice that the WG00 model constructed with a MW
extinction curve and a shell geometry produces a UV bump that is {\it
  stronger} than is observed in the MW extinction curve.  This follows
from the observationally--motivated assumption that the UV bump is a
true absorption feature.  In contrast, the dusty geometry produces
bump strengths both stronger and weaker than the bump in the MW
extinction curve, depending on the overall opacity.  This serves to
highlight the fact that UV bumps can be either stronger or weaker than
the underlying extinction curve depending on the large--scale
star--dust geometry.  Notice also that the dusty geometries produce
attenuation curves that are functions of the total dust column density
in the sense that higher column densities produce grayer $V-$band
normalized attenuation curves.

In this work we will consider the variation in broadband colors of
galaxies as a function of inclination.  Based on our understanding of
disk-dominated galaxies, we expect the parameter $\tau_d$, which
characterizes the $V-$band opacity of the diffuse dust, to vary with
inclination.  The attenuation curve may also vary with inclination,
because the dust column density will vary with inclination
\citep[e.g.,][]{Varosi99, Witt00, Tuffs04}.  However, we do not expect
the dust attenuation suffered by young stars to depend on inclination,
as this is a local phenomenon.  We will therefore not allow the
parameter $\tau_{bc}$ to vary with galaxy inclination. For our main
results we will also fix the attenuation curve for the young component
to be a power--law with index $\delta=-1.0$.  Notice that there is no
UV bump associated with the birth cloud dust.


\section{Results}
\label{s:res}

\begin{figure*}[!t]
\begin{center}
\resizebox{5.4in}{!}{\includegraphics{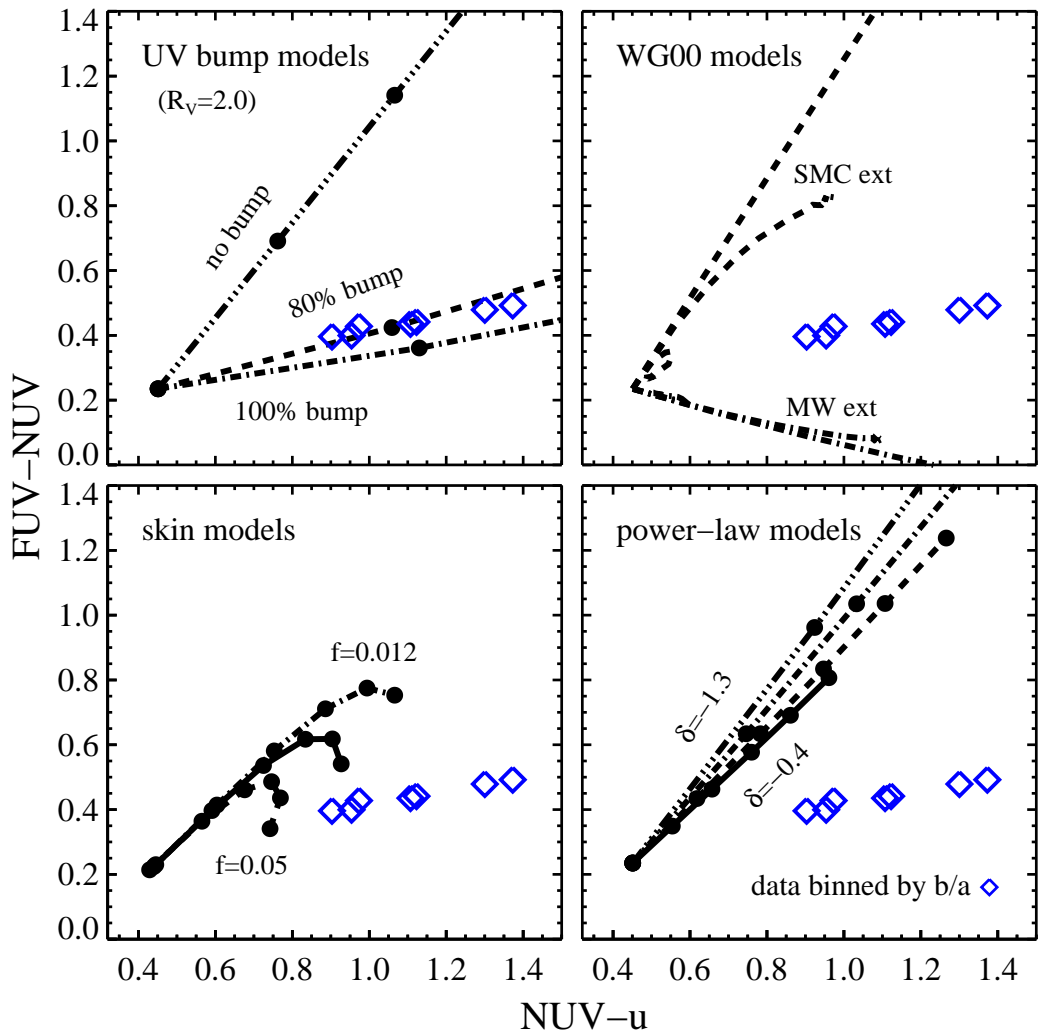}}
\end{center}
\vspace{0.7cm}
\caption{UV color--color diagrams comparing stellar population models
  to the average colors of low redshift disk--dominated galaxies as a
  function of inclination.  The data points become monotonically
  redder in NUV$-u$ with decreasing $b/a$ (increasing inclination).
  All models assume a constant SFH, are at solar metallicity, and
  adopt the same attenuation prescription for young stars.  The models
  differ only in the adopted dust attenuation curve for the diffuse
  dust.  For each attenuation curve except for those from WG00, the
  trajectory in color--color space is due to an increase in the
  diffuse dust opacity, $\tau_d$, from 0.0 to 1.5.  The opacity
  increases by intervals of 0.3, as indicated by the symbols along the
  models.  {\it Top left:} Milky Way attenuation curves with $R_V=2.0$
  and with different treatments of the UV bump at 2175\AA: standard MW
  bump strength, no bump, 80\% relative bump strength.  {\it Top
    right:} Attenuation curves based on the WG00 dust models.  Models
  are constructed with both SMC ({\it dashed lines}) and MW ({\it
    dot-dashed lines}) extinction curves.  With each extinction curve,
  attenuation curves were derived for three dust geometries (cloudy,
  dusty, and shell; see text).  {\it Bottom left:} Models constructed
  with a power-law attenuation curve ($\tau\propto\lambda^{-0.7}$),
  with a fraction of starlight that is completely unobscured.
  Fractions of $f=0.012$, $f=0.025$, and $f=0.050$ are shown.  {\it
    Bottom right:} Power--law attenuation curves ranging from
  $\lambda^{-0.4}$ to $\lambda^{-1.3}$ in steps of 0.3.}
\label{fig:incl}
\vspace{0.2cm}
\end{figure*}

\subsection{UV colors}
\label{s:uvcol}

In Figure \ref{fig:incl} we compare the average UV colors of a sample
of disk--dominated galaxies as a function of inclination to a suite of
models.  Recall that the data sample consists of galaxies at
$0.01<z<0.05$ with Sersic index $n<2.5$ and stellar masses in the
range $9.5<{\rm log}(M/M_\Sol)<10.0$.  Data are binned in inclination
with width 0.1, starting at $b/a=0.3$.  Within each bin the average
flux in each filter is computed.  The resulting colors redden
monotonically with increasing inclination.  Due to the large sample
size, the errors on the mean colors are in all cases smaller than the
symbol sizes.  We have computed errors both via bootstrap re-sampling
and the naive variance and find comparable results.

The model predictions all have the same underlying stellar population,
which consists of stars with solar metallicity having formed in a
constant mode of star formation from $t=0.0$ to $t=13.7$ Gyr.  For
each model starlight is attenuated according to a two--component dust
model, the details of which are described in $\S$\ref{s:dust}.  The
models differ principally in the attenuation curve associated with the
diffuse cirrus dust.  For each model, the sequence in colors is a
sequence in increasing $V-$band optical depth, $\tau_d$, associated
with this diffuse component.  The sequence increases from $\tau_d=0.0$
to $\tau_d=1.5$.

In the upper left panel of Figure \ref{fig:incl} we compare the sample
of disk--dominated galaxies to MW attenuation curves with $R_V=2.0$.
The curves differ only in the treatment of the UV bump at 2175\AA.  We
consider curves with a UV bump characteristic of the average MW
extinction \citep{Cardelli89}, no UV bump, and a UV bump with strength
equal to 80\% of the full value ($B=0.8$ in the notation of
$\S$\ref{s:attn}).  A model with a UV bump strength equal to 80\% of
the full value provides an excellent fit to the data, while a model
without a UV bump provides a poor match.  For this model, the
observations are reproduced for $0.2\lesssim\tau_d\lesssim0.5$.  This
range in $\tau_d$ is consistent with detailed radiative transfer
models that consider inclination-dependent attenuation
\citep{Tuffs04}.

While not shown, use of the Calzetti et al. attenuation curve for
starbursts produces results qualitatively similar to the no bump
curve, and therefore fails to reproduce the observed trends.  Also not
shown are models constructed with a MW curve with $R_V=3.1$ (the
average MW value).  These models also provide a poor match to the
data, both for no UV bump and a bump with full strength ($B=1.0$).  A
model with $R_V=3.1$ and $B=0.6$ provides a good fit to the UV data,
although the best-fit $\tau_d$ values as inferred from the UV colors
cannot reproduce the observed optical colors (see $\S$\ref{s:optcol}).
For this reason, the $R_V=3.1$ model is discarded.  In contrast, the
inclination-dependent UV and optical colors can be well-fit with the
same $\tau_d$ values for both the UV and optical colors.

We show model predictions for the attenuation curves of WG00 in the
upper right panel of Figure \ref{fig:incl}.  WG00 model predictions
are presented for the `homogeneous' local dust distribution, for all
three large--scale geometrical configurations (`cloudy', `dusty', and
`shell') utilizing both the MW and SMC extinction curves.  For these
models, unlike the others, the predictions are shown for optical
depths ranging from 0.0 to 50.0.  The WG00 models with MW extinction
all predict FUV-NUV colors too blue compared to the data.  With SMC
extinction the WG00 models qualitatively resemble the models with the
MW extinction curve without a UV bump shown in the upper left panel.
The `dusty' geometric configuration with SMC extinction is the model
that begins to bend toward redder NUV$-u$ colors.  For this geometry
the colors saturate because the attenuation curve becomes increasing
greyer at high opacity (see Figure \ref{fig:uvext}).  One therefore
cannot appeal to ever higher opacities to produce colors in better
agreement with the data.  The `clumpy' local dust distribution
predictions do not produce better fits to the data.  It is of course
possible that a different geometrical configuration not considered by
WG00 may yield better agreement with the data when the SMC extinction
curve is used.  Also, the spherical symmetry assumed in the WG00
models may be an important limitation for our purposes.  It is however
beyond the scope of this paper to consider such variations to the WG00
model.

In the lower right panel we show results for power--law attenuation
curves with an index that varies from $\delta=-0.4$ to $\delta=-1.3$.
These models all predict FUV-NUV colors far too red at a given NUV$-u$
color.

In the lower left panel we show results for a model constructed that
is identical to the power-law attenuation model with $\delta=-0.7$
except that a skin of completely unobscured stars is added with mass
fraction of $f=1-5$\%.  This model does somewhat better at describing
the data than the pure power--law models in that the dependence of the
FUV-NUV color on the NUV$-u$ color becomes shallow at large optical
depths.  The qualitative trends are a decent match to the data, but
the quantitative predictions are much too blue in NUV$-u$.  We will
come back to this point below.  Note also that the turnover in the
model predictions at high opacities is generic to skin models.  As the
opacity increases, eventually the bulk population becomes heavily
obscured and the (very blue) colors of the unobscured population
dominates the integrated color.  If the opacity were increased beyond
what is shown in the figure, the colors would eventually approach the
$\tau_d=0.0$ colors.

It is clear from Figure \ref{fig:incl} that none of the commonly used
attenuation curves provide a good description of the observed trends
with inclination.  This is so because the disk--dominated sample has
FUV-NUV colors that vary only slightly with inclination, while at the
same time NUV$-u$ varies by $\approx0.5$ mag.  The implication is that
the net attenuation in the FUV and NUV bands is comparable, i.e.,
$A_{\rm FUV}\approx A_{\rm NUV}$, and they both increase with
increasing inclination.  Most standard attenuation curves (e.g., a
power--law, the form advocated by Calzetti et al. for starburst
galaxies, a MW curve without a UV bump, and an SMC-based curve) do not
satisfy this requirement.  It is also difficult for geometrical
effects to produce such a trend.  The observed trend is however
readily explained if the attenuation curve contains a UV bump with
strength equal to 80\% of the canonical MW value, and with $R_V=2.0$.
In this case the net attenuation in the FUV and NUV bands is
comparable, as evidenced by the top left panel of Figure
\ref{fig:incl}.

\vspace{0.1cm}

\subsubsection{Assessing model assumptions}
\label{s:ass}

\begin{figure}[!t]
\begin{center}
\vspace{0.15cm}
\resizebox{3.3in}{!}{\includegraphics{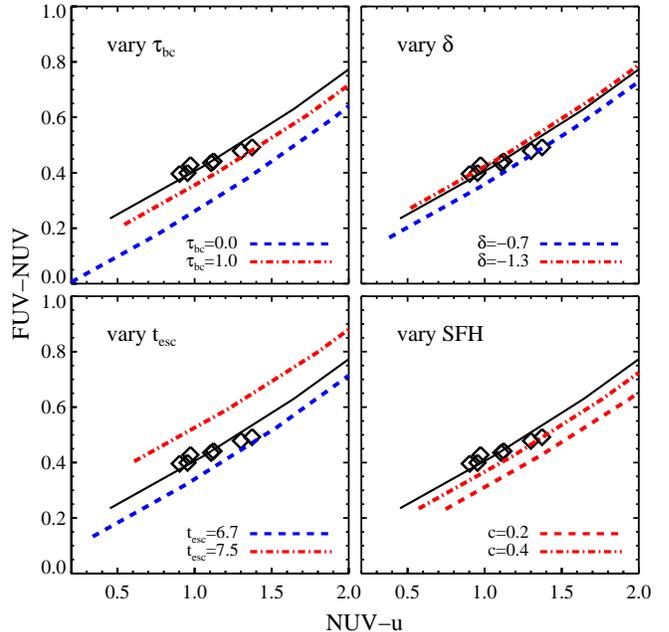}}
\end{center}
\vspace{0.2cm}
\caption{UV color--color diagram demonstrating the impact of various
  model assumptions.  The default model ({\it solid line}), which
  assumes MW attenuation with a UV bump ($B=0.8$) and $R_V=2.0$ and a
  constant SFH, and the observational results ({\it symbols}), are
  repeated in all panels. {\it Top left:} Variation in the $V-$band
  optical depth, $\tau_{bc}$ associated with young stars.  {\it Top
    right:} Variation in the index, $\delta$, of the attenuation curve
  for the young stars.  {\it Bottom left:} Variation in the time
  during which young stars experience additional attenuation.  This
  time is shown in the legend in units of log($t$/yr).  {\it Bottom
    right:} Variation in the SFH.  Models are shown where a smaller
  fraction of mass is formed in a constant mode of SFH ($c=0.4$ and
  $c=0.2$); the remaining mass is assumed to have formed instantly at
  $t=0.0$.  Notice the different axis scales compared to Figure
  \ref{fig:incl}.}
\label{fig:mtest}
\vspace{0.3cm}
\end{figure}

\begin{figure*}[!t]
\begin{center}
\resizebox{6.1in}{!}{\includegraphics{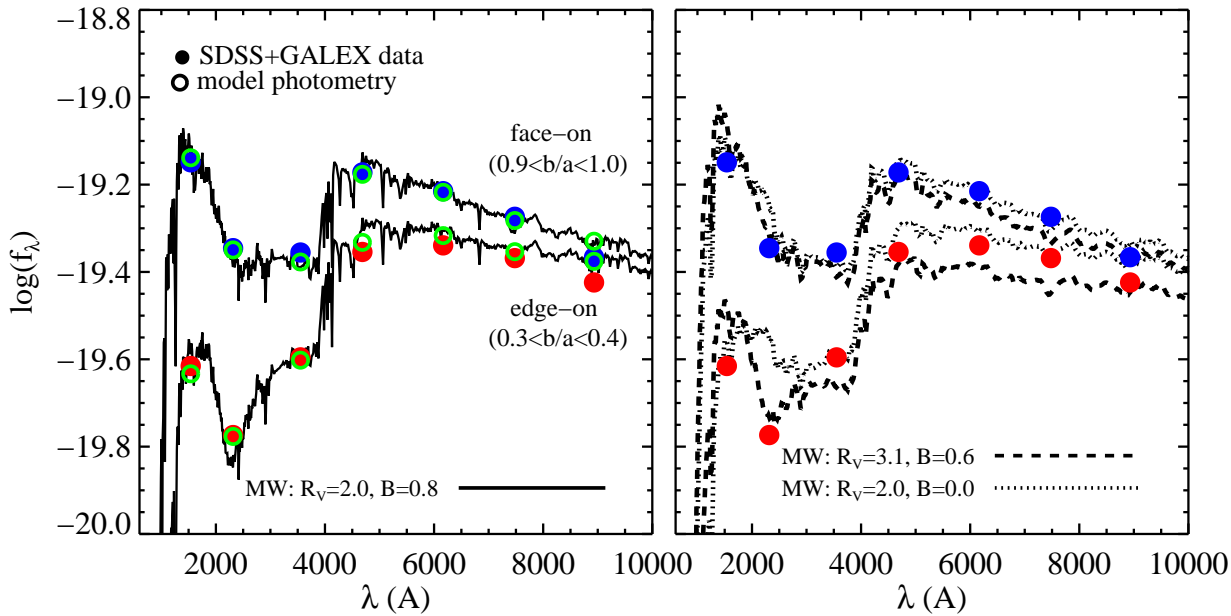}}
\end{center}
\vspace{0.3cm}
\caption{Average SEDs of face-on and edge-on star-forming galaxies
  with $9.5<{\rm log}(M/\Msun)<10.0$ ({\it solid symbols}).  Errors on
  the average SEDs are smaller than the symbol sizes.  These data are
  compared to models with a constant SFH, $Z=Z_\Sol$, and a MW-like
  attenuation curve.  {\it Left panel:} Our best-fit model attenuation
  curve with $R_V=2.0$ and $B=0.8$.  The open symbols indicate the
  corresponding model magnitudes computed from the model spectra.
  {\it Right panel:} Model attenuation curves with $R_V=2.0$, $B=0.8$
  and $R_V=3.1$, $B=0.6$.}
\label{fig:optspec}
\vspace{0.3cm}
\end{figure*}

In this section we consider the consequences of model variations
beyond the attenuation curve for diffuse dust.  In Figure
\ref{fig:mtest} we explore the impact of several important model
assumptions on the UV colors.  The default model here adopts the same
underlying stellar population as before (constant SFH, solar
metallicity), along with a MW attenuation curve with $R_V=2.0$ and
$B=0.8$, $\tau_{bc}=0.5$, a power--law attenuation with $\delta=-1.0$
for the young stars, and a transition time between young and old stars
of $t=10^7$ yr.  Variations to this default model that are shown in
Figure \ref{fig:mtest} include the following: no additional
attenuation around young stars ($\tau_{bc}=0.0$), twice the
attenuation around young stars ($\tau_{bc}=1.0$), an attenuation curve
for young stars with an index of $-0.7$ and $-1.3$, a transition time
between young and old stars, log($t_{\rm esc}$/yr), of 6.7 and 7.5,
and SFHs in which only a fraction of stars are formed in a constant
mode of star formation ($c=0.4$ and $c=0.2$), the rest having formed
instantly at $t=0.0$.

In all cases the modifications simply result in a shift in the model
locus in the FUV-NUV vs. NUV$-u$ plane.  Therefore, varying these
model parameters allows us the freedom to move the model predictions
in Figure \ref{fig:incl} horizontally and/or vertically by a few
tenths of a magnitude.  Such shifts do not change our conclusions
because the attenuation models that fail do so in their dependence of
FUV-NUV on NUV$-u$.  In other words, allowing the freedom to shift the
models arbitrarily does not produce significantly better fits to the
data, except for the skin models and the $B=1.0$ UV bump model; see
below.  Modifying the `zero-point' of the models will however clearly
change the quantitative conclusions drawn from this figure, such as
the average SFH and total dust opacity, but that is to be expected
when only considering UV colors.

The only cases where a shift in the NUV$-u$ color may result in a
better model fit is for the skin models and the $B=1.0$ bump model.
For the skin model, moving the model NUV$-u$ colors redward by
$\approx0.4$ mag would yield an improved agreement between the skin
models and the data.  Such a shift can be induced by a SFH where only
20\% of the mass forms in a constant mode ($c=0.2$), the rest having
formed instantly at $t=0.0$.  Below we show that such SFHs can be
ruled out by the optical colors of our disk--dominated sample.  For
the $B=1.0$ bump model, a shift redward in FUV-NUV of $\approx0.1$
mag, induced for example by choosing $t_{\rm esc}=10^{7.5}$ yr,
would yield a good match with the data.  We leave this as an open
possibility, noting that such a shift would only strengthen our
conclusion that a strong 2175\AA\, absorption feature was required to
match the observations.

\subsection{Constraints from optical colors}
\label{s:optcol}

We now turn to optical colors in order to gain additional insight into
the dust and stellar population content of our sample.  In Figure
\ref{fig:optspec} we show the average spectral energy distributions
(SEDs) for galaxies in our highest and lowest inclination bins.  SEDs
were constructed from {\it GALEX} and SDSS photometry.  We also
include in this figure our favored model, which has a constant SFH,
$Z=Z_\Sol$, and a MW dust attenuation curve with $R_V=2.0$ and $B=0.8$
(models with $R_V=2.0, B=0.0$ and $R_V=3.1, B=0.6$ are also shown for
comparison).  The models for the face-on and edge-on galaxies differ
only in the $V-$band optical depth associated with diffuse dust,
$\tau_d$.  Clearly our favored model matches the data well over the
interval $1500$\AA$\lesssim\lambda\lesssim9000$\AA, providing further
support for the assumptions made in our favored model.  The strong
depression in the SED of edge-on galaxies at $\lambda\approx2200$\AA\,
is striking, and strongly suggestive of the presence of the UV bump.

The disagreement between the models and data in the $z-$band is
noteworthy.  Unfortunately, no significant conclusions can be drawn
from this disagreement because the SED of star--forming galaxies at
$\lambda\gtrsim7000$\AA\, is sensitive to the presence of
thermally-pulsating asymptotic giant branch stars
\citep[e.g.,][]{Maraston05}.  The modeling of this stellar
evolutionary phase is very uncertain \citep[e.g.,][]{Conroy09a,
  Conroy10a}, and minor modifications to the bolometric luminosity and
temperature of these stars can readily account for the disagreement
seen in the figure.

\begin{figure}[!t]
\begin{center}
\vspace{0.22cm}
\resizebox{3.5in}{!}{\includegraphics{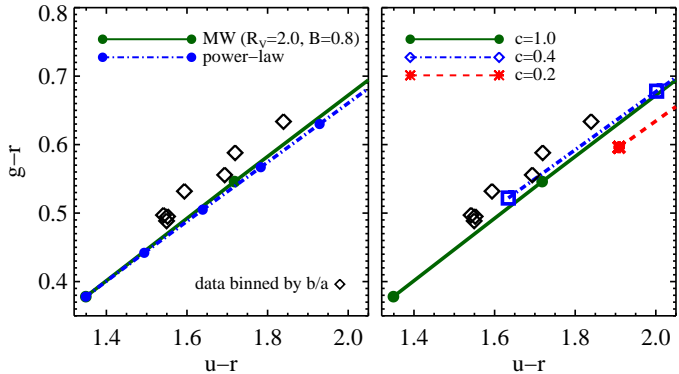}}
\end{center}
\vspace{0.2cm}
\caption{Optical color--color plots as a function of inclination
  compared to various models.  {\it Left panel:} Comparison of MW
  (with $R_V=2.0$ and $B=0.8$) and power-law (with $\delta=-0.7$)
  attenuation curves.  Here the SFH is constant ($c=1.0$).  Each model
  is a sequence in optical depth associated with diffuse dust, in
  steps of $\Delta(\tau_d)=0.3$. {\it Right panel:} Models with the
  same MW attenuation curve but varying SFH.  The SFHs are varied such
  that a fraction, $c$, of the mass is formed in a continuous mode of
  star formation, with the rest having formed instantly at $z=0.0$.}
\label{fig:optcol}
\vspace{0.3cm}
\end{figure}

In order to further assess our model assumptions, in Figure
\ref{fig:optcol} we consider optical color--color plots.  In this
figure the SDSS photometry is compared to models that differ in their
attenuation curve associated with diffuse dust (left panel), and
models that differ in their assumed SFH.  Note that the model $ugriz$
photometry is insensitive to the presence of a UV bump, and so we do
not distinguish between MW attenuation with or without a UV bump in
this figure.

The variations in SFH result in significant differences in these $ugr$
color-color plots.  Notice that the bluest points in both the $c=0.4$
and $c=0.2$ SFH models are {\it redder} than the bluest data points.
This implies that these SFHs cannot possibly fit our disk--dominated
sample (recall that a single SFH must be able to fit the colors for
all inclination bins, since SFH will not vary with inclination).  As
mentioned in $\S$\ref{s:data}, this is not surprising in light of
results from detailed modeling of the spectra of galaxies with masses
comparable to those considered herein \citep[e.g.,][]{Salim07,
  Brinchmann04}.  We note in passing that the SFH models considered
herein produce UV and optical colors that are almost indistinguishable
from $\tau-$model SFHs with $\tau=5$ and 8 Gyr corresponding to
$c=0.2$ and $c=0.4$.

As discussed in $\S$\ref{s:ass}, the skin model shown in the lower
left panel of Figure \ref{fig:incl} could be brought into agreement
with the data if a SFH with $c=0.2$ were adopted.  While not shown,
the optical colors for the skin model are identical to those in Figure
\ref{fig:optcol} for the MW curve because the flux contributed by the
skin is negligible in the optical.  The results from this figure
therefore allow us to confidently rule out the skin models because the
SFH necessary to reconcile the model with the UV colors grossly fails
to match the optical colors.

As mentioned in \ref{s:uvcol}, a model with $R_V=3.1$ and $B=0.6$ also
provided a good fit to the UV colors.  However, as shown in Figure
\ref{fig:optspec}, this model cannot simultaneously match the entire
UV through near-IR SEDs of star-forming galaxies.  The disagreement is
particularly striking for the NUV$-u$ colors, where the constraints on
low values for $R_V$ are strongest.

\subsection{Trends with stellar mass and metallicity, and variation in
  data analysis}
\label{s:trends}

\begin{figure}[!t]
\begin{center}
\resizebox{3.3in}{!}{\includegraphics{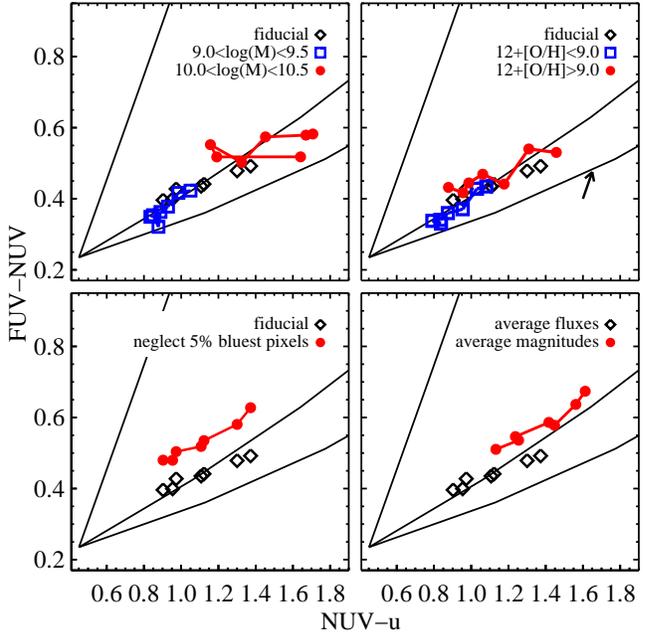}}
\end{center}
\vspace{0.3cm}
\caption{Color--color diagram comparing a variety of different
  selection criteria and methods of data analysis. Models with a MW
  attenuation curve with varying UV bump strength ($B=0.0,0.8,1.0$)
  are included for comparison ({\it solid lines}; these models are
  identical to the models shown in the upper left panel of Figure
  \ref{fig:incl}).  In each panel the fiducial sample is shown as
  diamonds.  Notice the different axis scales compared to Figure
  \ref{fig:incl}.  {\it Upper left:} Samples selected with stellar
  masses $9.0<{\rm log}(M/\Msun)<9.5$ and $10.0<{\rm
    log}(M/\Msun)<10.5$.  {\it Upper right:} Samples selected with
  12+[O/H]$<9.0$ and 12+[O/H]$>9.0$ drawn from a parent sample with
  $9.0<{\rm log}(M/\Msun)<10.0$.  The arrow indicates the change in
  colors accompanying a change in $Z$ of the underlying stellar
  population from $0.4Z_\Sol$ to $Z_\Sol$.  {\it Lower left:} FUV-NUV
  color computed after removing the 5\% bluest pixels.  {\it Lower
    right:} Comparison between colors of the fiducial sample computed
  by averaging magnitudes and averaging fluxes.}
\label{fig:svar}
\vspace{0.2cm}
\end{figure}

\begin{figure*}[!t]
\begin{center}
\resizebox{5.in}{!}{\includegraphics{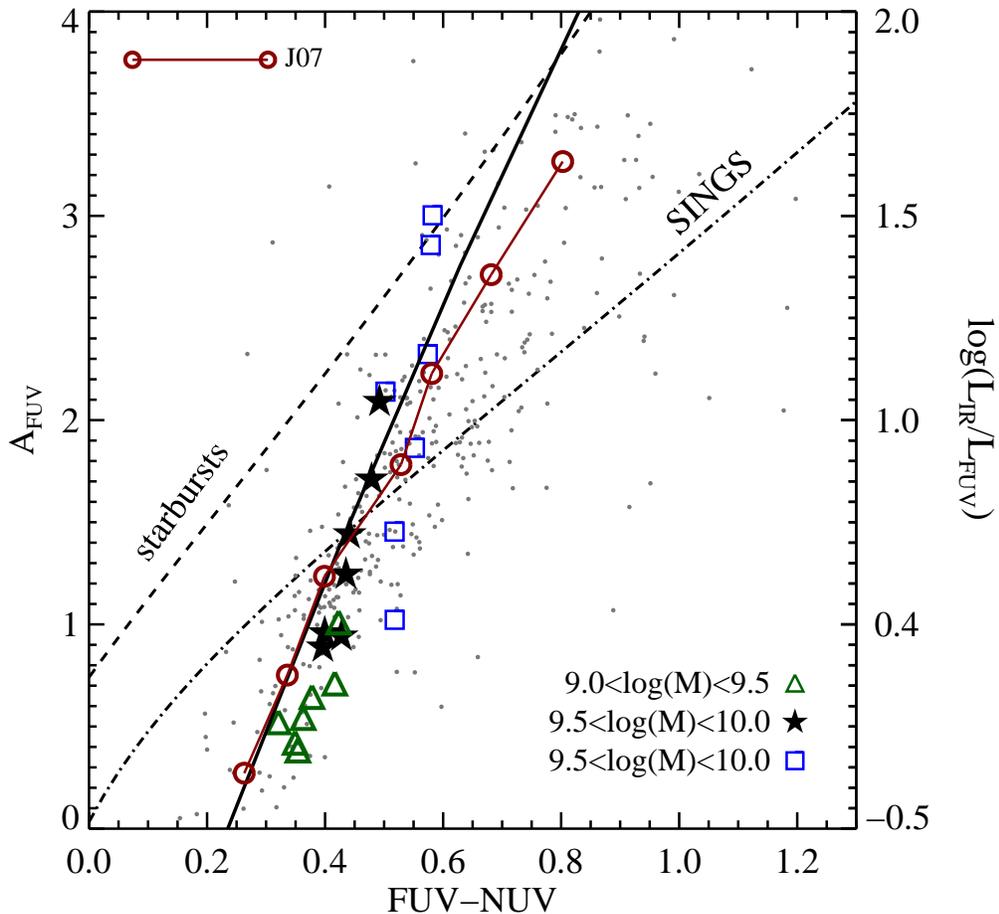}}
\end{center}
\caption{Total FUV attenuation as a function of FUV-NUV color (analog
  of the IRX--$\beta$ relation).  Lines show the relation for
  starburst galaxies \citep[\emph{dashed line};][]{Kong04}, for
  galaxies in the SINGS survey \citep[\emph{dot-dashed
    line};][]{Munoz-Mateos09}, and our best-fit model that contains a
  strong UV bump ({\it solid line}).  For the starburst and SINGS
  samples, $A_{\rm FUV}$ is estimated from $L_{\rm IR}/L_{\rm FUV}$
  according to the relation provided by \citet{Buat05}.  For our
  samples of disk--dominated galaxies ({\it diamond, star, and square
    symbols}), $A_{\rm FUV}$ is estimated as described in the text.
  We show results for three bins in stellar mass, as indicated in the
  legend.  Small symbols are from the work of
  \citet[][J07]{Johnson07b}; larger connected symbols are median
  FUV-NUV colors from J07 computed in bins of $A_{\rm FUV}$.}
\label{fig:uvx}
\vspace{0.2cm}
\end{figure*}

We now return to the UV colors.  In Figure \ref{fig:svar} we explore
the sensitivity of the results shown in Figure \ref{fig:incl} to the
sample selection and method of data analysis.  In this figure we
consider galaxies both more and less massive than our fiducial sample,
and galaxies split according to their gas--phase metallicity.  In the
latter case we consider a wider bin in mass, $9.0<{\rm
  log}(M/M_\Sol)<10.0$, in order to increase statistics.  In the panel
showing the dependence on metallicity, we also show the change in
model colors accompanying a change in metallicity of the underlying
stellar population from $0.4Z_\Sol$ to $Z_\Sol$.  Model predictions
for the default set of assumptions are included in this figure, where
we have adopted a MW attenuation curve with $R_V=2.0$ and with UV bump
strengths $B=0.0,0.8,1.0$.  These model predictions are identical to
those shown in the upper left panel of Figure \ref{fig:incl}.

The observed trends are generally as expected: more massive and more
metal--rich galaxies are redder in both NUV$-u$ and FUV-NUV, although
the differences are not large.  There is tentative evidence that the
more massive galaxies deviate from our favored model in the sense that
their FUV-NUV colors are $\approx0.1$ mag bluer than expected for
edge-on galaxies.  While we are hesitant to interpret this trend, if
real it could indicate a stronger UV bump in more massive galaxies.

The observed weak dependence on metallicity is generally consistent
with being due to the underlying stellar population, under the
assumption that variation in gas-phase metallicity is accompanied by a
similar change in the stellar metallicity.

In Figure \ref{fig:svar} we also show the observed trend where we have
computed the FUV-NUV color after masking the bluest 5\% of pixels.
This masking was done to test the extent to which the observed trend
(namely, the weak trend of FUV-NUV vs. inclination) is due to a few
very blue and concentrated regions of star formation.  The NUV$-u$
color was not re-computed with this masking because of the
complications in translating the masks in the {\it GALEX} pixels into
the SDSS pixels.  The resulting FUV-NUV colors are on average
$\approx0.1$ mag redder than the colors computed from all of the
pixels.  The results from this exercise demonstrate that the observed
very blue FUV-NUV colors even for the most inclined galaxies cannot be
attributed to a few very blue concentrated star--forming regions.

Finally, in this same figure we show how the average colors of our
fiducial sample depends on the method of averaging.  We show average
colors computed by averaging fluxes (our standard approach) and by
averaging magnitudes.  As mentioned in $\S$\ref{s:stack}, in the
absence of dust, averaging fluxes is the correct approach.  In the
presence of dust, there is no clearly preferred method, although we
note that the attenuation by dust is averaged in a more meaningful way
when averaging magnitudes.  Regardless, the resulting average
inclination-dependent colors are largely unaffected by our choice of
averaging.

\subsection{The IRX$-\beta$ relation}

Recall that the IRX$-\beta$ relation compares the UV spectral slope,
$\beta$, to the ratio of total infrared to UV luminosity, the latter
being a fairly direct probe of the total UV attenuation.  In theory,
an IRX$-\beta$ relation therefore provides a simple estimate of the
net attenuation in the UV based solely on the UV slope (i.e., the
FUV-NUV color).  In practice, this procedure is complicated not only
by the large observed scatter in the relation, but also by the fact
that starburst and `normal' star--forming galaxies follow different
relations \citep{Meurer99, Dale07, Boissier07}, and the fact that the
relation depends on SFH \citep{Kong04, Cortese08}.

In Figure \ref{fig:uvx} we translate our results into an analog of the
IRX$-\beta$ relation.  Specifically, we estimate the attenuation in
the FUV, $A_{\rm FUV}$, of the observed galaxies from their $g-r$
color by simply comparing the model colors to the observed colors in
order to estimate the total net attenuation [i.e., $A_{\rm
  FUV}\propto{\rm ln}(F_i^{\rm FUV}/F_o^{\rm FUV})$ where $i$ and $o$
denote the dust--free and dust reddened model spectra, respectively].
We obtain similar results when using other colors to estimate $A_{\rm
  FUV}$.  We then shift our estimated $A_{\rm FUV}$ values down by 0.8
mag in order to match the locus of previous results (see below).  The
particular values of $A_{\rm FUV}$ will clearly depend on our model
assumptions (e.g., the assumed SFH), but the general behavior of our
results in the IRX$-\beta$ should be robust.  Results are shown for
three stellar mass bins.

In this figure we include the IRX$-\beta$ relation estimated for a
sample of local starburst galaxies \citep{Meurer99, Kong04}, and
normal star--forming galaxies from the SINGS survey
\citep{Munoz-Mateos09}.  The `IRX' portion of these relations has been
converted into $A_{\rm FUV}$ via the formula provided in
\citet{Buat05}.

The most striking conclusion to be drawn from Figure \ref{fig:uvx} is
that our sample spans a large range in $A_{\rm FUV}$ but a
comparatively narrow range in FUV-NUV color, i.e., our results imply
an extremely steep IRX$-\beta$ relation.  The relation for our sample
of disk--dominated galaxies is much steeper than either the starburst
or normal star--forming galaxy relation.  This result is driven
entirely by the fact that while FUV-NUV varies little in our sample,
other colors, such as NUV$-u$, $u-r$, and $g-r$ vary significantly.
The trend for the highest stellar mass bin is notable in that it
appears to be steeper than our favored model.  As mentioned in
$\S$\ref{s:trends}, if real this trend could point to a stronger UV
bump in more massive galaxies.

The IRX$-\beta$ relation from the SINGS sample is representative of
the entire SINGS sample, excluding only three starburst galaxies and
galaxies with FUV-NUV>0.9 \citep{Munoz-Mateos09}.  It is therefore not
surprising that the SINGS relation is much shallower than ours,
because their sample contains galaxies with a wide range in SFHs.
Indeed, from Figure 3 in \citet{Munoz-Mateos09} it is apparent that
their IRX$-\beta$ relation is driven largely by variations in SFH, in
the sense that galaxies with redder FUV-NUV colors have formed the
bulk of their stars at earlier epochs. In contrast, the IRX$-\beta$
relation derived herein is characteristic of a single SFH (for each
stellar mass bin), and therefore our relation is truly a sequence in
dust opacity.

In this figure we also compare our results to
\citet[][J07]{Johnson07b} who computed the IRX$-\beta$ relation for
$\approx$1000 galaxies using observations from {\it GALEX} and {\it
  Spitzer}\footnote{The data from J07 shown in Figure \ref{fig:uvx}
  are revised quantities based on the same sample as in J07.  The most
  significant change with respect to the data presented in J07 is the
  use of updated {\it GALEX} photometry, which results in bluer
  FUV-NUV colors by $\approx0.05-0.1$ mag.}.  We show their results
for all galaxies with $D_n4000<1.5$, where $D_n4000$ is a measure of
the strength of the 4000\AA\, break. $D_n4000$ is a measure of the age
of the stellar population that is relatively insensitive to dust.
This cut on $D_n4000$ selects actively star-forming galaxies, similar
to those in our own sample.  For this sample we also show the median
FUV-NUV color in bins of $A_{\rm FUV}$, where the FUV attenuation has
been estimated from IRX using the formula in \citet{Buat05}.

The average relation from J07 is in excellent agreement with our
observational results and also with our favored model that includes a
strong UV bump in the diffuse dust attenuation curve, after we shift
our results downward in Figure \ref{fig:uvx} by 0.8 mag.  We stress
that the solid line in Figure \ref{fig:uvx} is {\it not} a fit to the
average relation from J07.  Rather, the agreement is remarkable
independent confirmation of our best-fit model.  We will discuss the
implications of this result in the following section.

The required shift of our results 0.8 mag downward to match the
results from J07 deserves further comment.  Note first that the
results from J07 and those derived herein are {\it not} directly
comparable, as J07 measure IRX while we only measure UV through
near-IR SEDs.  Any attempted comparison between our work and J07 will
therefore be model-dependent.  A detailed analysis of this issue is
beyond the scope of the present article.  However, irrespective of our
results, notice that the implied $A_{\rm FUV}$ vs. FUV$-$NUV relation
in the J07 sample extrapolates to $A_{\rm FUV}=0.0$ at a FUV$-$NUV
color significantly redder than canonical dust-free solar metallicity
stellar population predictions (FUV$-$NUV$\approx0$).  This indicates
that there may be some modeling error in translating IRX (which J07
actually measure) into $A_{\rm FUV}$ \citep[which was derived from IRX
via the relation from][]{Buat05}, at least at low attenuation.

\begin{figure}[!t]
\resizebox{3.5in}{!}{\includegraphics{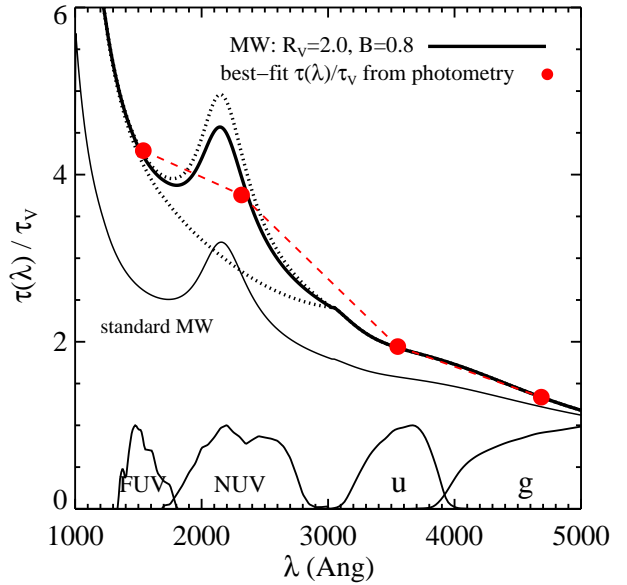}}
\caption{Attenuation curves normalized to the $V-$band.  The dotted
  curves show MW attenuation curves with $R_V=2.0$, both for a
  standard UV bump strength and no UV bump ($B=1.0$ and $B=0.0$,
  respectively).  The thick solid line is our best--fit attenuation
  curve with $R_V=2.0$ and a UV bump strength equal to 80\% of the
  standard MW value.  The thin solid line is the standard MW curve
  (i.e., with $R_V=3.1$ and $B=1.0$). The points show the actual
  constraints on the attenuation curve provided by the broadband
  photometry. Transmission curves with arbitrary normalization are
  included for the FUV, NUV, $u$, and $g$ filters.}
\label{fig:best}
\vspace{0.3cm}
\end{figure}


\section{Discussion}
\label{s:disc}

\subsection{An emerging physical picture, and implications}

In the previous section we demonstrated that the
inclination--dependent UV colors of a mass--selected sample of
disk--dominated galaxies are best explained by a MW attenuation curve
with $R_V=2.0$ that includes a prominent UV bump at 2175\AA.  The
strength of this bump is approximately 80\% as strong as the average
UV bump observed in the MW.  This result is summarized in Figure
\ref{fig:best}.  In this figure we show our favored attenuation curve
and compare this to the standard MW curve ($R_V=3.1$).  We also show
the actual constraints on the attenuation provided by the FUV, NUV,
$u$, and $g$ filters.  It is clear from this figure that while our
results are consistent with a strong UV bump, we cannot claim that our
results {\it require} a UV bump, since we rely solely on broadband
photometry over a narrow redshift range.  However, inspection of
Figure \ref{fig:optspec} strongly suggests that the UV bump is
responsible, as we know of no other model variation capable of
reproducing the depression in the average SED of edge-on galaxies at
$\lambda\approx2200$\AA.

Our results imply that previous work attempting to measure the SFR in
low redshift galaxies has underestimated the amount of attenuation in
the near-UV.  From inspection of Figure \ref{fig:best} we can see that
typical attenuation in the NUV band is underestimated by approximately
the value of the $V-$band optical depth, $\tau_V$, when comparing
standard attenuation curves (power-law, starburst, and average MW) to
our favored model.  For moderately dusty galaxies the underestimation
may therefore be substantial.  The bias is such that SFRs based
primarily on the near-UV will be lower than the intrinsic SFR.
Additional work will be required to understand in detail how the
effects of a strong UV bump propagate into the derived physical
properties of galaxies.

From this figure we can also understand qualitatively why our
disk--dominated sample favors an attenuation curve with $R_V=2.0$, as
opposed to the canonical value of $R_V=3.1$ preferred in the
MW.\footnote{Of course, as emphasized throughout, attenuation and
  extinction are conceptually different, and so there is no {\it a
    priori} reason for the attenuation curve that best describes our
  sample to be similar to the average extinction curve measured for
  the MW.}  Decreasing $R_V$ from 3.1 to 2.0 results in substantially
more attenuation between the $u$ and NUV filters, and therefore a much
redder NUV$-u$ color.  The relative attenuation between FUV and NUV,
and also between $u$ and $g$ is does not change substantially between
the $R_V=3.1$ and $R_V=2.0$ curves, and so these colors change by much
less than NUV$-u$.  Therefore, it is principally the constraint from
the NUV$-u$ color that drives the requirement for $R_V\approx2.0$.

If our interpretation of the observed trends is correct, this would
constitute the first detection of the UV bump in the attenuation
properties of galaxies at low redshift.  The only other detection of
the UV bump in the attenuation curves of galaxies was reported by
\citet{Noll09}, who considered restframe UV spectra of star-forming
galaxies at $z\sim2$.  In \citet{Conroy10b} the observed $B-R$ colors
of star-forming galaxies at $0.7<z<1.4$ were used to constrain the
average strength of the UV bump.  A bump as strong as observed in the
MW was ruled out for $R_V=3.1$, as we find herein, but a UV bump
strength needed to match our sample ($B\approx0.8$ with $R_V=2.0$,
which, in the UV, is comparable to $B\approx0.6$ with $R_V=3.1$) could
not be ruled out.  More detailed modeling of the $z\sim1$ population
will be required to place stronger constraints on the presence of the
UV bump.  Luckily, the necessary data already exists.

It is important to understand why evidence of the UV bump in the
attenuation curve of galaxies has proved elusive until now.  Many
authors have previously attempted to model the UV colors of galaxies
with a standard MW attenuation curve (i.e., with $R_V=3.1$ and
$B=1.0$).  In all cases the MW curve was found to be a poor fit, as we
find herein.  The reduction in the strength of the UV bump to $B=0.8$,
accompanied by our choice of $R_V=2.0$, is essential for our
attenuation models to fit the data.  Recall again that an attenuation
curve with $R_V=2.0$ and $B=0.8$ yields FUV-NUV colors comparable to
an attenuation curve with $R_V=3.1$ and $B=0.6$.  In other words, if
considering only the UV and $R_V=3.1$ MW-like attenuation curves, a
significant depression of the UV bump is required.  Clearly, ruling
out the standard MW curve does not imply that observed galaxies show
no evidence for a UV bump.  In addition, considering the UV colors of
galaxies as a function of inclination allowed us to confidently
separate stellar population effects from dust attenuation effects.

These results must somehow be reconciled with the well--known fact
that local starburst galaxies show no evidence for a UV bump
\citep[e.g.,][]{Calzetti94}.  It has been conjectured that either
metallicity, intensity of the interstellar radiation field, or the
star--dust geometry modulates the strength of the UV bump.  The dust
model of \citet{Silva98}, as explored in \citet{Granato00}, provides a
plausible path toward reconciling our results with those for the
starburst galaxies.  Granato et al. showed that an underlying MW
extinction curve (applied to dust in the diffuse ISM) can give rise to
a starburst attenuation curve similar to that observed by Calzetti et
al., and can also give rise to an attenuation curve with a significant
UV bump for `normal' star-forming galaxies.  In their model the
difference is due to geometry: for the starbursts the UV dust
attenuation is dominated by dust within the molecular clouds in which
the young stars are embedded, because the fraction of young stars is
high in starbursts.  The molecular clouds are optically thick, and so
the resulting attenuation curve is governed primarily by the
wavelength-dependent fraction of the intrinsic flux emitted by young
stars.  In the case of normal star-forming galaxies, a significant
fraction of UV photons are emitted by stars that have left their birth
clouds, and so the only attenuation they suffer is due to the
optically thin diffuse ISM, which may imprint a UV bump.

More recently, \citep{Panuzzo07} have used the Silva et al. dust models
to interpret the IRX$-\beta$ relation for normal SF galaxies and
conclude that the data are consistent with a UV bump in the extinction
curve.  These authors highlight the important role of age-dependent
extinction on the resulting attenuation curve.  \citet{Inoue06}
consider the effects of dust scattering in detail and conclude that
the data are consistent with a UV bump.  Finally, \citet{Burgarella05}
take an empirical approach in modeling UV attenuation similar to what
we do herein, and conclude that the observed IRX$-\beta$ relation for
normal SF galaxies is consistent with a UV bump with half the strength
of the UV bump in the MW extinction curve.  All of these studies
provide additional support to our principle conclusion regarding the
presence of a UV bump in the attenuation curve of normal SF galaxies.

While the Silva et al. model is capable of explaining the absence of a
UV bump in starburst galaxies and its presence in normal star-forming
galaxies by geometrical effects, other possibilities remain.  The most
plausible alternative explanation is the preferential destruction of
the grains responsible for the UV bump in the harsh interstellar
radiation characteristic of starburst galaxies.  PAH emission in the
IR is known to be modulated by the strength of the radiation field in
the sense that harder radiation fields result in diminished (or even
negligible) PAH emission \citep[e.g.,][]{Voit92, Madden06, Smith07}.
These properties of PAHs can provide a natural explanation for the
variation in UV bump strength, if PAHs are responsible for the
absorption at 2175\AA.  More directly, the UV bump is absent in the
extinction curve probed by four out of five sightlines in the SMC
\citep{Pei92}, which may be due to the harsher radiation field there
\citep{Gordon03}.  Metallicity is an unlikely explanation for the
difference between starburst and normal galaxies because the gas-phase
metallicities of these two types of galaxies are not substantially
different \citep{Calzetti94, Tremonti04}.  Now that we have identified
samples of galaxies that plausibly contain a UV bump, each of these
scenarios can be tested.

In Figure \ref{fig:uvx}, we showed that the average relation between
total UV attenuation and UV spectral slope is extremely steep in the
sense that a narrow range in observed spectral slope (i.e., FUV-NUV
color) corresponds to a large range in UV attenuation.  These results
extend and strengthen previous results that also found very steep
relations \citep[e.g.,][]{Bell02b, Johnson07a}.  In our
interpretation, the steepness of this relation is due to the
properties of the average attenuation curve for a fixed average SFH.
In essence, the attenuation curve flattens considerably over the
wavelength range probed by the FUV and NUV filters
($1500$\AA$\lesssim\lambda\lesssim2300$\AA), and so large variation in
the total attenuation produces little change in FUV-NUV colors.
Appeal to the UV bump at 2175\AA\, provides a natural explanation for
the observed flattening in the attenuation curve.  The shallower
IRX$-\beta$ relations found in previous work arises from the fact that
previous relations were constructed from samples of galaxies spanning
a wide range of SFHs.

We have presented {\it average} UV colors as a function of inclination
and UV attenuation.  For individual galaxies, one can expect
variations about these average relations due to variations in either
the star-dust geometry, SFH, metallicity, or dust-to-gas ratio.
Indeed, within a given inclination bin the spread in FUV-NUV colors is
large ($\sigma\approx0.5$ mag).  We speculate that variation in the SFH
over the last $\sim10^8$ yr and variation in the star-dust geometry
are the two largest sources of scatter.  In addition, we clearly
detect systematic changes in the FUV-NUV color with stellar mass,
which may either be due to a varying attenuation curve or possibly a
varying average SFH with galaxy mass.

In light of the steepness of the IRX$-\beta$ relation for a given SFH
as measured herein, its dependence on the SFH, and the expected
galaxy-to-galaxy scatter due to variations in other physical
properties, we conclude that the uncertainties in estimating UV
attenuation based on the observed UV slope are very large and believe
that this calls the utility of the IRX$-\beta$ relation into question.
This conclusion applies when $\beta$ --- the UV slope --- is measured
over a wavelength interval including the UV bump.  If this spectral
range is avoided then the IRX$-\beta$ relation may yield more reliable
results.  Care must clearly be taken when estimating $\beta$ from
either spectroscopy or photometry.

\begin{figure}[!t]
\resizebox{3.5in}{!}{\includegraphics{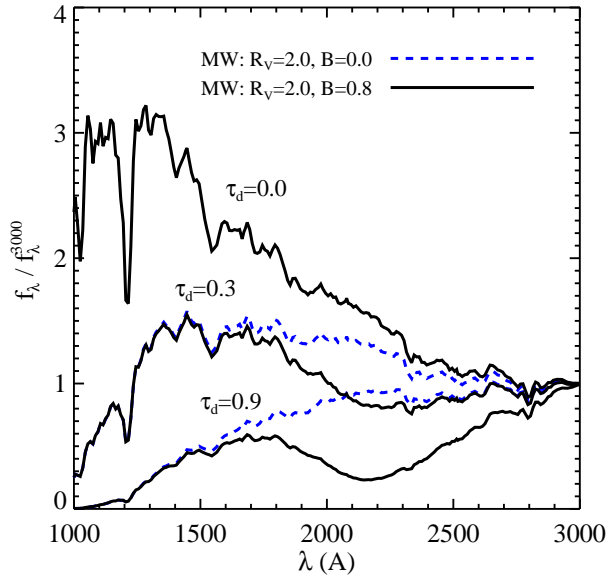}}
\caption{Restframe UV spectra of two models that differ only in the
  strength of the UV bump at 2175\AA.  One model has no UV bump while
  the other has a UV bump with strength 80\% of the average MW
  strength.  Each model is a sequence in optical depth characterizing
  the diffuse dust component: $\tau_d=0.0,0.3,0.9$.  The spectra are
  normalized to the flux at 3000\AA. }
\label{fig:uvspec}
\vspace{0.2cm}
\end{figure}

\subsection{Future directions}

The results contained in this work present clear directions for future
work.  First, moderate resolution spectra in the UV would provide
conclusive evidence for the UV bump.  In Figure \ref{fig:uvspec} we
show the restframe UV spectra for our default stellar population model
(constant SFH and solar metallicity), for two dust attenuation curves
that differ only in the strength of the UV bump.  We show models both
for a bump strength of 80\% (the favored dust model) and zero bump
strength.  Spectra are shown for three values of the $V-$band optical
depth characterizing the diffuse dust: $\tau_d=0.0,0.3,0.9$.  The
treatment of dust surrounding young stars is constant and the same as
previous sections.  For reference, the average inclination-dependent
colors of our disk--dominated sample can be reproduced in our favored
model for $0.2\lesssim\tau_d\lesssim0.5$.

This figure demonstrates that UV spectra of moderately dusty galaxies
would provide a clear test of our preferred model.  Our sample of
highly inclined disk--dominated star-forming galaxies would be ideal
for these purposes.  Narrow band photometry around 2175\AA\, could
also potentially reveal the unique signature of the UV bump.  The
required filters are already on-board the {\it Hubble Space
  Telescope}.

If the UV bump is indeed responsible for the observed trends, the next
task will be to identify whether or not PAHs are responsible for the
absorption.  Recall that the identity of the grain(s) responsible for
the UV bump is still a mystery, and, while PAHs are a plausible
candidate, other possibilities remain \citep[including, e.g.,
graphite;][]{Draine93}.  If PAHs are responsible, then one would
expect to uncover a strong correlation between PAH emission in the
infrared \citep[traced for example by the $8\mu m$ channel on the {\it
  Spitzer Space Telescope;}][]{Munoz-Mateos09} and a strong UV bump
(traced for example by the FUV-NUV color).

An especially promising direction would be to consider galaxies with
gas-phase metallicities both above and below 12+[O/H]$=8.1$.  Based on
data from the SINGS survey, \citet{Draine07} found that galaxies above
this metallicity contained a wide range of PAH emission strengths,
whereas below this value galaxies were deficient in PAH emission.  We
would therefore expect the trends of FUV-NUV with inclination to
change qualitatively below this metallicity scale, again assuming that
PAHs are the cause of the UV bump.  According to \citet{Tremonti04},
galaxies with such low gas-phase metallicities have stellar masses
log$(M/\Msun)\lesssim 8.5$.  Irregular galaxies are more common than
disk--dominated galaxies at this mass scale, and so it may be
difficult to separate trends with geometry from trends with PAH
abundance across this mass scale.  Nonetheless, with detailed modeling
the PAH hypothesis may be tested with the approach outlined above.


\section{Summary}

In this work, we measure the UV and optical colors of disk--dominated
galaxies at $0.01<z<0.05$ as a function of their inclination.  These
inclination-dependent colors provide powerful constraints on dust
attenuation models because the observed trends cannot be caused by
variations in the underlying stellar populations.  The observed
inclination-dependent FUV-NUV and NUV$-u$ colors cannot be explained
by common dust attenuation curves, including power-law models,
standard MW and SMC curves, nor the attenuation curve measured for
starbursts \citep{Calzetti94}.  After considering a number of model
variations, we are led to conclude that the strong dust feature at
2175\AA\, (the UV bump) seen in the MW extinction curve is responsible
for the observed trends.  In order to simultaneously fit the UV and
optical colors, we require an attenuation curve analogous to the MW
extinction curve with $R_V\equiv A_V/E(B-V)\approx2.0$ and a UV bump
strength equal to 80\% of the standard MW value.  Previous work failed
to recognize the importance of the UV bump because only standard MW
curves were considered ($R_V=3.1$, full UV bump strength), and these
consistently fail to match observed trends.

We also construct an analog of the IRX$-\beta$ relation and find that
the inferred relation between UV slope and FUV attenuation is
extremely steep.  Our derived relation agrees well with recent work
that combines UV and IR photometry for $\approx$1000 low redshift
galaxies \citep{Johnson07b}, but is much steeper than the starburst
relation.  Our results provide a natural interpretation of the
IRX$-\beta$ relations found in previous work.  Namely, for a given
SFH, the IRX$-\beta$ relation will be very steep because of the
properties of the attenuation curve.  Varying the SFH will produce
primarily horizontal shifts in this relation.  When considering
heterogeneous samples of galaxies with a range of SFHs, the
IRX$-\beta$ relation will on average appear shallow, due to the range
in SFHs, but will show tremendous scatter, due to the attenuation
curve.  From the steepness of this relation for a given SFH, we
conclude that it is not possible to accurately correct for dust
attenuation --- even in an average sense --- based solely on the UV
slope.

Confirmation of our interpretation must await UV spectroscopic
follow-up.  If confirmed, these results will pave the way toward
finally identifying the grain population responsible for the 2175\AA\,
extinction feature.  In addition, these results necessitate a more
nuanced interpretation of the UV spectral slopes of star-forming
galaxies.


\acknowledgments 

CC is supported by the Porter Ogden Jacobus Fellowship at Princeton
University.  We thank the MPA/JHU collaboration for making their
catalogs publically available, and Bruce Draine for comments on an
earlier draft.  We also thank Ben Johnson for providing an updated
version of his data used in our Figure 8.

Funding for the Sloan Digital Sky Survey (SDSS) has been provided by
the Alfred P. Sloan Foundation, the Participating Institutions, the
National Aeronautics and Space Administration, the National Science
Foundation, the U.S. Department of Energy, the Japanese
Monbukagakusho, and the Max Planck Society. The SDSS Web site is
http://www.sdss.org/.

The SDSS is managed by the Astrophysical Research Consortium (ARC) for
the Participating Institutions. The Participating Institutions are The
University of Chicago, Fermilab, the Institute for Advanced Study, the
Japan Participation Group, The Johns Hopkins University, Los Alamos
National Laboratory, the Max-Planck-Institute for Astronomy (MPIA),
the Max-Planck-Institute for Astrophysics (MPA), New Mexico State
University, University of Pittsburgh, Princeton University, the United
States Naval Observatory, and the University of Washington.

This work made extensive use of the NASA Astrophysics
Data System and of the {\tt astro-ph} preprint archive at {\tt
  arXiv.org}.


\begin{thebibliography}{111}
\expandafter\ifx\csname natexlab\endcsname\relax\def\natexlab#1{#1}\fi

\bibitem[{{Adelman-McCarthy} {et~al.}(2008)}]{DR6}
{Adelman-McCarthy}, J.~K. {et~al.} 2008, \apjs, 175, 297

\bibitem[{{Bell}(2002)}]{Bell02b}
{Bell}, E.~F. 2002, \apj, 577, 150

\bibitem[{{Bianchi} {et~al.}(1996){Bianchi}, {Clayton}, {Bohlin}, {Hutchings},
  \& {Massey}}]{Bianchi96}
{Bianchi}, L., {Clayton}, G.~C., {Bohlin}, R.~C., {Hutchings}, J.~B., \&
  {Massey}, P. 1996, \apj, 471, 203

\bibitem[{{Blanton} \& {Roweis}(2007)}]{Blanton07}
{Blanton}, M.~R. \& {Roweis}, S. 2007, \aj, 133, 734

\bibitem[{{Blanton} {et~al.}(2005)}]{Blanton05}
{Blanton}, M.~R. {et~al.} 2005, \aj, 129, 2562

\bibitem[{{Boissier} {et~al.}(2004){Boissier}, {Boselli}, {Buat}, {Donas}, \&
  {Milliard}}]{Boissier04}
{Boissier}, S., {Boselli}, A., {Buat}, V., {Donas}, J., \& {Milliard}, B. 2004,
  \aap, 424, 465

\bibitem[{{Boissier} {et~al.}(2007)}]{Boissier07}
{Boissier}, S. {et~al.} 2007, \apjs, 173, 524

\bibitem[{{Bouwens} {et~al.}(2009){Bouwens}, {Illingworth}, {Franx}, {Chary},
  {Meurer}, {Conselice}, {Ford}, {Giavalisco}, \& {van Dokkum}}]{Bouwens09}
{Bouwens}, R.~J., {Illingworth}, G.~D., {Franx}, M., {Chary}, R., {Meurer},
  G.~R., {Conselice}, C.~J., {Ford}, H., {Giavalisco}, M., \& {van Dokkum}, P.
  2009, \apj, 705, 936

\bibitem[{{Brinchmann} {et~al.}(2004){Brinchmann}, {Charlot}, {White},
  {Tremonti}, {Kauffmann}, {Heckman}, \& {Brinkmann}}]{Brinchmann04}
{Brinchmann}, J., {Charlot}, S., {White}, S.~D.~M., {Tremonti}, C.,
  {Kauffmann}, G., {Heckman}, T., \& {Brinkmann}, J. 2004, \mnras, 351, 1151

\bibitem[{{Buat} {et~al.}(2005)}]{Buat05}
{Buat}, V. {et~al.} 2005, \apjl, 619, L51

\bibitem[{{Bundy} {et~al.}(2009)}]{Bundy09}
{Bundy}, K. {et~al.} 2009, ArXiv:0912.1077

\bibitem[{{Burgarella} {et~al.}(2005){Burgarella}, {Buat}, \&
  {Iglesias-P{\'a}ramo}}]{Burgarella05}
{Burgarella}, D., {Buat}, V., \& {Iglesias-P{\'a}ramo}, J. 2005, \mnras, 360,
  1413

\bibitem[{{Calzetti}(2001)}]{Calzetti01}
{Calzetti}, D. 2001, \pasp, 113, 1449

\bibitem[{{Calzetti} {et~al.}(2000){Calzetti}, {Armus}, {Bohlin}, {Kinney},
  {Koornneef}, \& {Storchi-Bergmann}}]{Calzetti00}
{Calzetti}, D., {Armus}, L., {Bohlin}, R.~C., {Kinney}, A.~L., {Koornneef}, J.,
  \& {Storchi-Bergmann}, T. 2000, \apj, 533, 682

\bibitem[{{Calzetti} {et~al.}(1994){Calzetti}, {Kinney}, \&
  {Storchi-Bergmann}}]{Calzetti94}
{Calzetti}, D., {Kinney}, A.~L., \& {Storchi-Bergmann}, T. 1994, \apj, 429, 582

\bibitem[{{Calzetti} {et~al.}(2005)}]{Calzetti05}
{Calzetti}, D. {et~al.} 2005, \apj, 633, 871

\bibitem[{{Capak} {et~al.}(2009)}]{Capak10}
{Capak}, P. {et~al.} 2009, ArXiv:0910.0444

\bibitem[{{Cardelli} {et~al.}(1989){Cardelli}, {Clayton}, \&
  {Mathis}}]{Cardelli89}
{Cardelli}, J.~A., {Clayton}, G.~C., \& {Mathis}, J.~S. 1989, \apj, 345, 245

\bibitem[{{Chabrier}(2003)}]{Chabrier03}
{Chabrier}, G. 2003, \pasp, 115, 763

\bibitem[{{Charlot} \& {Fall}(2000)}]{Charlot00}
{Charlot}, S. \& {Fall}, S.~M. 2000, \apj, 539, 718

\bibitem[{{Conroy}(2010)}]{Conroy10b}
{Conroy}, C. 2010, \mnras, 404, 247

\bibitem[{{Conroy} \& {Gunn}(2010)}]{Conroy10c}
{Conroy}, C. \& {Gunn}, J.~E. 2010, \apj, 712, 833

\bibitem[{{Conroy} {et~al.}(2009){Conroy}, {Gunn}, \& {White}}]{Conroy09a}
{Conroy}, C., {Gunn}, J.~E., \& {White}, M. 2009, \apj, 699, 486

\bibitem[{{Conroy} {et~al.}(2010){Conroy}, {White}, \& {Gunn}}]{Conroy10a}
{Conroy}, C., {White}, M., \& {Gunn}, J.~E. 2010, \apj, 708, 58

\bibitem[{{Cortese} {et~al.}(2008){Cortese}, {Boselli}, {Franzetti}, {Decarli},
  {Gavazzi}, {Boissier}, \& {Buat}}]{Cortese08}
{Cortese}, L., {Boselli}, A., {Franzetti}, P., {Decarli}, R., {Gavazzi}, G.,
  {Boissier}, S., \& {Buat}, V. 2008, \mnras, 386, 1157

\bibitem[{{da Cunha} {et~al.}(2008){da Cunha}, {Charlot}, \&
  {Elbaz}}]{daCunha08}
{da Cunha}, E., {Charlot}, S., \& {Elbaz}, D. 2008, \mnras, 388, 1595

\bibitem[{{da Cunha} {et~al.}(2010){da Cunha}, {Eminian}, {Charlot}, \&
  {Blaizot}}]{daCunha10}
{da Cunha}, E., {Eminian}, C., {Charlot}, S., \& {Blaizot}, J. 2010, \mnras,
  403, 1894

\bibitem[{{Dale} {et~al.}(2007)}]{Dale07}
{Dale}, D.~A. {et~al.} 2007, \apj, 655, 863

\bibitem[{{Disney} {et~al.}(1989){Disney}, {Davies}, \& {Phillipps}}]{Disney89}
{Disney}, M., {Davies}, J., \& {Phillipps}, S. 1989, \mnras, 239, 939

\bibitem[{{Draine}(2003)}]{Draine03}
{Draine}, B.~T. 2003, \araa, 41, 241

\bibitem[{{Draine}(2009)}]{Draine09}
---. 2009, ArXiv:0903.1658

\bibitem[{{Draine} \& {Malhotra}(1993)}]{Draine93}
{Draine}, B.~T. \& {Malhotra}, S. 1993, \apj, 414, 632

\bibitem[{{Draine} {et~al.}(2007)}]{Draine07}
{Draine}, B.~T. {et~al.} 2007, \apj, 663, 866

\bibitem[{{Driver} {et~al.}(2007){Driver}, {Popescu}, {Tuffs}, {Liske},
  {Graham}, {Allen}, \& {de Propris}}]{Driver07}
{Driver}, S.~P., {Popescu}, C.~C., {Tuffs}, R.~J., {Liske}, J., {Graham},
  A.~W., {Allen}, P.~D., \& {de Propris}, R. 2007, \mnras, 379, 1022

\bibitem[{{Elbaz} {et~al.}(1999)}]{Elbaz99}
{Elbaz}, D. {et~al.} 1999, \aap, 351, L37

\bibitem[{{El{\'{\i}}asd{\'o}ttir} {et~al.}(2009)}]{Ardis09}
{El{\'{\i}}asd{\'o}ttir}, {\'A}. {et~al.} 2009, \apj, 697, 1725

\bibitem[{{Engelbracht} {et~al.}(2005){Engelbracht}, {Gordon}, {Rieke},
  {Werner}, {Dale}, \& {Latter}}]{Engelbracht05}
{Engelbracht}, C.~W., {Gordon}, K.~D., {Rieke}, G.~H., {Werner}, M.~W., {Dale},
  D.~A., \& {Latter}, W.~B. 2005, \apjl, 628, L29

\bibitem[{{Galliano} {et~al.}(2008){Galliano}, {Dwek}, \&
  {Chanial}}]{Galliano08}
{Galliano}, F., {Dwek}, E., \& {Chanial}, P. 2008, \apj, 672, 214

\bibitem[{{Gil de Paz} {et~al.}(2007)}]{GildePaz07}
{Gil de Paz}, A. {et~al.} 2007, \apj, 661, 115

\bibitem[{{Giovanelli} {et~al.}(1994){Giovanelli}, {Haynes}, {Salzer},
  {Wegner}, {da Costa}, \& {Freudling}}]{Giovanelli94}
{Giovanelli}, R., {Haynes}, M.~P., {Salzer}, J.~J., {Wegner}, G., {da Costa},
  L.~N., \& {Freudling}, W. 1994, \aj, 107, 2036

\bibitem[{{Giovanelli} {et~al.}(1995){Giovanelli}, {Haynes}, {Salzer},
  {Wegner}, {da Costa}, \& {Freudling}}]{Giovanelli95}
---. 1995, \aj, 110, 1059

\bibitem[{{Gordon} {et~al.}(2003){Gordon}, {Clayton}, {Misselt}, {Landolt}, \&
  {Wolff}}]{Gordon03}
{Gordon}, K.~D., {Clayton}, G.~C., {Misselt}, K.~A., {Landolt}, A.~U., \&
  {Wolff}, M.~J. 2003, \apj, 594, 279

\bibitem[{{Goudfrooij} {et~al.}(1994){Goudfrooij}, {Hansen}, {Jorgensen}, \&
  {Norgaard-Nielsen}}]{Goudfrooij94}
{Goudfrooij}, P., {Hansen}, L., {Jorgensen}, H.~E., \& {Norgaard-Nielsen},
  H.~U. 1994, \aaps, 105, 341

\bibitem[{{Granato} {et~al.}(2000){Granato}, {Lacey}, {Silva}, {Bressan},
  {Baugh}, {Cole}, \& {Frenk}}]{Granato00}
{Granato}, G.~L., {Lacey}, C.~G., {Silva}, L., {Bressan}, A., {Baugh}, C.~M.,
  {Cole}, S., \& {Frenk}, C.~S. 2000, \apj, 542, 710

\bibitem[{{Inoue} {et~al.}(2006){Inoue}, {Buat}, {Burgarella}, {Panuzzo},
  {Takeuchi}, \& {Iglesias-P{\'a}ramo}}]{Inoue06}
{Inoue}, A.~K., {Buat}, V., {Burgarella}, D., {Panuzzo}, P., {Takeuchi}, T.~T.,
  \& {Iglesias-P{\'a}ramo}, J. 2006, \mnras, 370, 380

\bibitem[{{Issa} {et~al.}(1990){Issa}, {MacLaren}, \& {Wolfendale}}]{Issa90}
{Issa}, M.~R., {MacLaren}, I., \& {Wolfendale}, A.~W. 1990, \aap, 236, 237

\bibitem[{{Jarrett} {et~al.}(2000){Jarrett}, {Chester}, {Cutri}, {Schneider},
  {Skrutskie}, \& {Huchra}}]{Jarrett00}
{Jarrett}, T.~H., {Chester}, T., {Cutri}, R., {Schneider}, S., {Skrutskie}, M.,
  \& {Huchra}, J.~P. 2000, \aj, 119, 2498

\bibitem[{{Johnson} {et~al.}(2007{\natexlab{a}})}]{Johnson07b}
{Johnson}, B.~D. {et~al.} 2007{\natexlab{a}}, \apjs, 173, 377

\bibitem[{{Johnson} {et~al.}(2007{\natexlab{b}})}]{Johnson07a}
---. 2007{\natexlab{b}}, \apjs, 173, 392

\bibitem[{{Kauffmann} {et~al.}(2003)}]{Kauffmann03a}
{Kauffmann}, G. {et~al.} 2003, \mnras, 341, 33

\bibitem[{{Knapp} {et~al.}(1989){Knapp}, {Guhathakurta}, {Kim}, \&
  {Jura}}]{Knapp89}
{Knapp}, G.~R., {Guhathakurta}, P., {Kim}, D., \& {Jura}, M.~A. 1989, \apjs,
  70, 329

\bibitem[{{Kong} {et~al.}(2004){Kong}, {Charlot}, {Brinchmann}, \&
  {Fall}}]{Kong04}
{Kong}, X., {Charlot}, S., {Brinchmann}, J., \& {Fall}, S.~M. 2004, \mnras,
  349, 769

\bibitem[{{Leger} \& {Puget}(1984)}]{Leger84}
{Leger}, A. \& {Puget}, J.~L. 1984, \aap, 137, L5

\bibitem[{{Leitherer} {et~al.}(1999)}]{Leitherer99}
{Leitherer}, C. {et~al.} 1999, \apjs, 123, 3

\bibitem[{{Lupton} {et~al.}(2001){Lupton}, {Gunn}, {Ivezi{\'c}}, {Knapp}, \&
  {Kent}}]{Lupton01}
{Lupton}, R., {Gunn}, J.~E., {Ivezi{\'c}}, Z., {Knapp}, G.~R., \& {Kent}, S.
  2001, in Astronomical Society of the Pacific Conference Series, Vol. 238,
  Astronomical Data Analysis Software and Systems X, ed. {F.~R.~Harnden Jr.,
  F.~A.~Primini, \& H.~E.~Payne}, 269

\bibitem[{{Madden} {et~al.}(2006){Madden}, {Galliano}, {Jones}, \&
  {Sauvage}}]{Madden06}
{Madden}, S.~C., {Galliano}, F., {Jones}, A.~P., \& {Sauvage}, M. 2006, \aap,
  446, 877

\bibitem[{{Maller} {et~al.}(2009){Maller}, {Berlind}, {Blanton}, \&
  {Hogg}}]{Maller09}
{Maller}, A.~H., {Berlind}, A.~A., {Blanton}, M.~R., \& {Hogg}, D.~W. 2009,
  \apj, 691, 394

\bibitem[{{Maraston}(2005)}]{Maraston05}
{Maraston}, C. 2005, \mnras, 362, 799

\bibitem[{{Martin} {et~al.}(2005)}]{Martin05}
{Martin}, D.~C. {et~al.} 2005, \apjl, 619, L1

\bibitem[{{Masters} {et~al.}(2003){Masters}, {Giovanelli}, \&
  {Haynes}}]{Masters03}
{Masters}, K.~L., {Giovanelli}, R., \& {Haynes}, M.~P. 2003, \aj, 126, 158

\bibitem[{{Masters} {et~al.}(2010)}]{Masters10}
{Masters}, K.~L. {et~al.} 2010, \mnras, 404, 792

\bibitem[{{Mediavilla} {et~al.}(2005){Mediavilla}, {Mu{\~n}oz}, {Kochanek},
  {Falco}, {Arribas}, \& {Motta}}]{Mediavilla05}
{Mediavilla}, E., {Mu{\~n}oz}, J.~A., {Kochanek}, C.~S., {Falco}, E.~E.,
  {Arribas}, S., \& {Motta}, V. 2005, \apj, 619, 749

\bibitem[{{M{\'e}nard} {et~al.}(2009){M{\'e}nard}, {Scranton}, {Fukugita}, \&
  {Richards}}]{Menard09}
{M{\'e}nard}, B., {Scranton}, R., {Fukugita}, M., \& {Richards}, G. 2009,
  ArXiv:0902.4240

\bibitem[{{Meurer} {et~al.}(1999){Meurer}, {Heckman}, \& {Calzetti}}]{Meurer99}
{Meurer}, G.~R., {Heckman}, T.~M., \& {Calzetti}, D. 1999, \apj, 521, 64

\bibitem[{{Morrissey} {et~al.}(2007)}]{Morrissey07}
{Morrissey}, P. {et~al.} 2007, \apjs, 173, 682

\bibitem[{{Motta} {et~al.}(2002){Motta}, {Mediavilla}, {Mu{\~n}oz}, {Falco},
  {Kochanek}, {Arribas}, {Garc{\'{\i}}a-Lorenzo}, {Oscoz}, \&
  {Serra-Ricart}}]{Motta02}
{Motta}, V., {Mediavilla}, E., {Mu{\~n}oz}, J.~A., {Falco}, E., {Kochanek},
  C.~S., {Arribas}, S., {Garc{\'{\i}}a-Lorenzo}, B., {Oscoz}, A., \&
  {Serra-Ricart}, M. 2002, \apj, 574, 719

\bibitem[{{Mu{\~n}oz-Mateos} {et~al.}(2009)}]{Munoz-Mateos09}
{Mu{\~n}oz-Mateos}, J.~C. {et~al.} 2009, \apj, 701, 1965

\bibitem[{{Natta} \& {Panagia}(1984)}]{Natta84}
{Natta}, A. \& {Panagia}, N. 1984, \apj, 287, 228

\bibitem[{{Noll} {et~al.}(2009)}]{Noll09}
{Noll}, S. {et~al.} 2009, \aap, 499, 69

\bibitem[{{O'Dowd} {et~al.}(2009)}]{ODowd09}
{O'Dowd}, M.~J. {et~al.} 2009, \apj, 705, 885

\bibitem[{{Oke} \& {Gunn}(1983)}]{Oke83}
{Oke}, J.~B. \& {Gunn}, J.~E. 1983, \apj, 266, 713

\bibitem[{{Panuzzo} {et~al.}(2007){Panuzzo}, {Granato}, {Buat}, {Inoue},
  {Silva}, {Iglesias-P{\'a}ramo}, \& {Bressan}}]{Panuzzo07}
{Panuzzo}, P., {Granato}, G.~L., {Buat}, V., {Inoue}, A.~K., {Silva}, L.,
  {Iglesias-P{\'a}ramo}, J., \& {Bressan}, A. 2007, \mnras, 375, 640

\bibitem[{{Pei}(1992)}]{Pei92}
{Pei}, Y.~C. 1992, \apj, 395, 130

\bibitem[{{Pierini} {et~al.}(2004){Pierini}, {Gordon}, {Witt}, \&
  {Madsen}}]{Pierini04}
{Pierini}, D., {Gordon}, K.~D., {Witt}, A.~N., \& {Madsen}, G.~J. 2004, \apj,
  617, 1022

\bibitem[{{Reddy} {et~al.}(2010){Reddy}, {Erb}, {Pettini}, {Steidel}, \&
  {Shapley}}]{Reddy10}
{Reddy}, N.~A., {Erb}, D.~K., {Pettini}, M., {Steidel}, C.~C., \& {Shapley},
  A.~E. 2010, \apj, 712, 1070

\bibitem[{{Reddy} {et~al.}(2006){Reddy}, {Steidel}, {Fadda}, {Yan}, {Pettini},
  {Shapley}, {Erb}, \& {Adelberger}}]{Reddy06a}
{Reddy}, N.~A., {Steidel}, C.~C., {Fadda}, D., {Yan}, L., {Pettini}, M.,
  {Shapley}, A.~E., {Erb}, D.~K., \& {Adelberger}, K.~L. 2006, \apj, 644, 792

\bibitem[{{Salim} {et~al.}(2007)}]{Salim07}
{Salim}, S. {et~al.} 2007, \apjs, 173, 267

\bibitem[{{Salim} {et~al.}(2009)}]{Salim09}
---. 2009, \apj, 700, 161

\bibitem[{{Schlegel} {et~al.}(1998){Schlegel}, {Finkbeiner}, \&
  {Davis}}]{Schlegel98}
{Schlegel}, D.~J., {Finkbeiner}, D.~P., \& {Davis}, M. 1998, \apj, 500, 525

\bibitem[{{Sersic}(1968)}]{Sersic68}
{Sersic}, J.~L. 1968, {Atlas de galaxias australes} (Cordoba, Argentina:
  Observatorio Astronomico, 1968)

\bibitem[{{Silva} {et~al.}(1998){Silva}, {Granato}, {Bressan}, \&
  {Danese}}]{Silva98}
{Silva}, L., {Granato}, G.~L., {Bressan}, A., \& {Danese}, L. 1998, \apj, 509,
  103

\bibitem[{{Smith} {et~al.}(2007)}]{Smith07}
{Smith}, J.~D.~T. {et~al.} 2007, \apj, 656, 770

\bibitem[{{Stecher}(1965)}]{Stecher65}
{Stecher}, T.~P. 1965, \apj, 142, 1683

\bibitem[{{Stone}(1991)}]{Stone91}
{Stone}, R.~C. 1991, \aj, 102, 333

\bibitem[{{Stratta} {et~al.}(2007){Stratta}, {Maiolino}, {Fiore}, \&
  {D'Elia}}]{Stratta07}
{Stratta}, G., {Maiolino}, R., {Fiore}, F., \& {D'Elia}, V. 2007, \apjl, 661,
  L9

\bibitem[{{Takeuchi} {et~al.}(2010)}]{Takeuchi10}
{Takeuchi}, T.~T. {et~al.} 2010, \aap, 514, A4+

\bibitem[{{Thilker} {et~al.}(2007)}]{Thilker07}
{Thilker}, D.~A. {et~al.} 2007, \apjs, 173, 538

\bibitem[{{Tojeiro} {et~al.}(2009){Tojeiro}, {Wilkins}, {Heavens}, {Panter}, \&
  {Jimenez}}]{Tojeiro09}
{Tojeiro}, R., {Wilkins}, S., {Heavens}, A.~F., {Panter}, B., \& {Jimenez}, R.
  2009, \apjs, 185, 1

\bibitem[{{Tremonti} {et~al.}(2004)}]{Tremonti04}
{Tremonti}, C.~A. {et~al.} 2004, \apj, 613, 898

\bibitem[{{Trumpler}(1930)}]{Trumpler30}
{Trumpler}, R.~J. 1930, \pasp, 42, 214

\bibitem[{{Tuffs} {et~al.}(2004){Tuffs}, {Popescu}, {V{\"o}lk}, {Kylafis}, \&
  {Dopita}}]{Tuffs04}
{Tuffs}, R.~J., {Popescu}, C.~C., {V{\"o}lk}, H.~J., {Kylafis}, N.~D., \&
  {Dopita}, M.~A. 2004, \aap, 419, 821

\bibitem[{{Unterborn} \& {Ryden}(2008)}]{Unterborn08}
{Unterborn}, C.~T. \& {Ryden}, B.~S. 2008, \apj, 687, 976

\bibitem[{{V{\'a}rosi} \& {Dwek}(1999)}]{Varosi99}
{V{\'a}rosi}, F. \& {Dwek}, E. 1999, \apj, 523, 265

\bibitem[{{V{\'a}zquez} \& {Leitherer}(2005)}]{Vazquez05}
{V{\'a}zquez}, G.~A. \& {Leitherer}, C. 2005, \apj, 621, 695

\bibitem[{{Voit}(1992)}]{Voit92}
{Voit}, G.~M. 1992, \mnras, 258, 841

\bibitem[{{Wang} \& {Heckman}(1996)}]{Wang96}
{Wang}, B. \& {Heckman}, T.~M. 1996, \apj, 457, 645

\bibitem[{{Wang} {et~al.}(2004){Wang}, {Hall}, {Ge}, {Li}, \&
  {Schneider}}]{Wang04}
{Wang}, J., {Hall}, P.~B., {Ge}, J., {Li}, A., \& {Schneider}, D.~P. 2004,
  \apj, 609, 589

\bibitem[{{Wang} {et~al.}(2008)}]{Wang08}
{Wang}, R. {et~al.} 2008, \apj, 687, 848

\bibitem[{{Weingartner} \& {Draine}(2001)}]{Weingartner01}
{Weingartner}, J.~C. \& {Draine}, B.~T. 2001, \apj, 548, 296

\bibitem[{{Werk} {et~al.}(2010)}]{Werk10}
{Werk}, J.~K. {et~al.} 2010, \aj, 139, 279

\bibitem[{{Williams} {et~al.}(2009){Williams}, {Quadri}, {Franx}, {van Dokkum},
  \& {Labb{\'e}}}]{Williams09}
{Williams}, R.~J., {Quadri}, R.~F., {Franx}, M., {van Dokkum}, P., \&
  {Labb{\'e}}, I. 2009, \apj, 691, 1879

\bibitem[{{Witt} \& {Gordon}(1996)}]{Witt96}
{Witt}, A.~N. \& {Gordon}, K.~D. 1996, \apj, 463, 681

\bibitem[{{Witt} \& {Gordon}(2000)}]{Witt00}
---. 2000, \apj, 528, 799

\bibitem[{{Witt} {et~al.}(1992){Witt}, {Thronson}, \& {Capuano}}]{Witt92}
{Witt}, A.~N., {Thronson}, Jr., H.~A., \& {Capuano}, Jr., J.~M. 1992, \apj,
  393, 611

\bibitem[{{Xilouris} {et~al.}(1999){Xilouris}, {Byun}, {Kylafis}, {Paleologou},
  \& {Papamastorakis}}]{Xilouris99}
{Xilouris}, E.~M., {Byun}, Y.~I., {Kylafis}, N.~D., {Paleologou}, E.~V., \&
  {Papamastorakis}, J. 1999, \aap, 344, 868

\bibitem[{{Xu} \& {Buat}(1995)}]{Xu95}
{Xu}, C. \& {Buat}, V. 1995, \aap, 293, L65

\bibitem[{{Yip} {et~al.}(2010){Yip}, {Szalay}, {Wyse}, {Dobos}, {Budav{\'a}ri},
  \& {Csabai}}]{Yip10}
{Yip}, C., {Szalay}, A.~S., {Wyse}, R.~F.~G., {Dobos}, L., {Budav{\'a}ri}, T.,
  \& {Csabai}, I. 2010, \apj, 709, 780

\bibitem[{{York} {et~al.}(2000)}]{York00}
{York}, D.~G. {et~al.} 2000, \aj, 120, 1579

\bibitem[{{York} {et~al.}(2006)}]{York06}
---. 2006, \mnras, 367, 945

\bibitem[{{Zaritsky}(1994)}]{Zaritsky94b}
{Zaritsky}, D. 1994, \aj, 108, 1619

\bibitem[{{Zubko} {et~al.}(2004){Zubko}, {Dwek}, \& {Arendt}}]{Zubko04}
{Zubko}, V., {Dwek}, E., \& {Arendt}, R.~G. 2004, \apjs, 152, 211

\end{thebibliography}


\begin{appendix}

  \section{The Milky Way Extinction Curve for Arbitrary UV Bump
    Strength}

  In this Appendix, we provide formulae to construct an extinction
  curve analogous to the average MW extinction curve, but for
  arbitrary UV bump strength.  We adopt the \citet{Cardelli89}
  parameterization of the MW extinction and only modify their formulae
  where necessary.  Below, $B$ will be the single parameter
  characterizing the strength of the UV bump at 2175\AA.  A value of
  $B=0.0$ results in no UV bump, while a value of $B=1.0$ returns the
  standard MW UV bump strength.  An IDL routine that computes the
  extinction curve for arbitrary $R_V$ and $B$ is available upon
  request from the authors.

  Cardelli et al. parameterize the MW extinction curve in the
  following way:

\be
A(\lambda)/A_V = a(x) + b(x)/R,
\ee
\noindent
where $x$ has units of $\mu m^{-1}$, $R\equiv E(B-V)/A_V$,
$A(\lambda)$ is the wavelength-dependent extinction in magnitudes,
$A_V$ is the extinction in the $V-$band, and $a(x)$ and $b(x)$ are
specified in the following piece-wise continuous way:

In the infrared: $0.3\mu m^{-1}<x<1.1\mu m^{-1}$,
\bea
a(x) & = & 0.574 \,x^{1.61}, \\
b(x) & = & -0.527\,x^{1.61}.
\eea

In the optical/near--IR: $1.1\mu m^{-1}<x<3.3\mu m^{-1}$ and $y\equiv x-1.82$,
\bea
a(x) & = & 1+0.177\,y-0.504\,y^2-0.0243\,y^3+0.721\,y^4+0.0198\,y^5-
0.775\,y^6+0.330\,y^7,\\
b(x) & = & 1.413\,y+2.283\,y^2+1.072\,y^3-5.384\,y^4-0.622\,y^5+
5.303\,y^6-2.090\,y^7.
\eea

In the near/mid--UV: $3.3\mu m^{-1}<x<5.9\mu m^{-1}$,
\bea
f_a & = & \bigg(\frac{3.3}{x}\bigg)^6\,
(-0.0370 + 0.0469\,B-0.601\,B/R + 0.542/R),\\
a(x) & = & 1.752-0.316\,x-\frac{0.104\,B}{(x-4.67)^2+0.341}+f_a,\\
b(x) & = & -3.09+1.825\,x+\frac{1.206\,B}{(x-4.62)^2+0.263}.
\eea

In the far--UV: $5.9\mu m^{-1}<x<8.0\mu m^{-1}$,
\bea
f_a &=& -0.0447\,(x-5.9)^2-0.00978\,(x-5.9)^3,\\
f_b &=&\,\,\,\,\, 0.213\,(x-5.9)^2+0.121\,(x-5.9)^3,\\
a(x) &=& 1.752-0.316\,x-\frac{0.104\,B}{(x-4.67)^2+0.341}+f_a,\\
b(x) &=& -3.09+1.825\,x+\frac{1.206\,B}{(x-4.62)^2+0.263}+f_b.
\eea

\end{appendix}

\end{document}